\documentclass[11pt, twocolumn]{aastex631}
\usepackage{hyperref}
\usepackage{multirow}

\begin{document}
	
\title{
\textbf{
\large{
Resolved Stellar Mass Estimation of Nearby Late-type Galaxies for the SPHEREx Era: 
Dependence on Stellar Population Synthesis Models
}}}

\author[0000-0003-3301-759X]{Jeong Hwan Lee}
\affil{Research Institute of Basic Sciences, Seoul National University, Seoul 08826, Republic of Korea}
\affil{Department of Physics and Astronomy, Seoul National University, 1 Gwanak-ro, Gwanak-gu, Seoul 08826, Republic of Korea}
\affil{The Center for High Energy Physics, Kyungpook National University, Daegu 41566, Republic of Korea}
\affil{Department of Astronomy and Atmospheric Sciences, Kyungpook National University, Daegu 41566, Republic of Korea}
\author[0000-0002-3560-0781]{Minjin Kim}
\affil{Department of Astronomy and Atmospheric Sciences, Kyungpook National University, Daegu 41566, Republic of Korea}
\author[0000-0002-5857-5136]{Taehyun Kim}
\affil{Department of Astronomy and Atmospheric Sciences, Kyungpook National University, Daegu 41566, Republic of Korea}
\author[0000-0002-4179-2628]{Hyunjin Shim}
\affil{Department of Earth Science Education, Kyungpook National University, Daegu 41566, Republic of Korea}
\author[0000-0001-6947-5846]{Luis C. Ho}
\affil{Kavli Institute for Astronomy and Astrophysics, Peking University, Beijing 100871, People’s Republic of China}
\affil{Department of Astronomy, School of Physics, Peking University, Beijing 100871, People’s Republic of China}
\author[0000-0003-3428-7612]{Ho Seong Hwang}
\affil{Department of Physics and Astronomy, Seoul National University, 1 Gwanak-ro, Gwanak-gu, Seoul 08826, Republic of Korea}
\affil{SNU Astronomy Research Center, Seoul National University, Seoul 08826, Republic of Korea}
\affil{Australian Astronomical Optics - Macquarie University, 105 Delhi Road, North Ryde, NSW 2113, Australia}
\author[0000-0002-4179-2628]{Hyunmi Song}
\affil{Department of Astronomy and Space Science, Chungnam National University, Daejeon 34134, Republic of Korea}
\author[0000-0002-6925-4821]{Dohyeong Kim}
\affil{Department of Earth Sciences, Pusan National University, Busan 46241, Republic of Korea}
\author[0000-0003-3078-2763]{Yujin Yang}
\affil{Korea Astronomy and Space Science Institute, Daejeon 34055, Republic of Korea}
\affil{University of Science and Technology, Korea, Daejeon 34113, Republic of Korea}
\author[0000-0002-2770-808X]{Woong-Seob Jeong}
\affil{Korea Astronomy and Space Science Institute, Daejeon 34055, Republic of Korea}
\affil{University of Science and Technology, Korea, Daejeon 34113, Republic of Korea}

\correspondingauthor{Minjin Kim}
\email{mkim.astro@gmail.com}

\begin{abstract}

The upcoming all-sky infrared spectrophotometric SPHEREx mission is set to provide spatially resolved stellar mass maps of nearby galaxies, offering more detailed insights than integrated light observations.
In this study, we develop a strategy for estimating stellar mass using SPHEREx by examining the dependence on different stellar population synthesis (SPS) models and proposing new scaling relations based on simulated SPHEREx data.
We estimate the resolved stellar masses of 19 nearby late-type galaxies from the PHANGS-MUSE survey, treating these as fiducial masses.
By testing four SPS models covering infrared wavelengths, i.e., E-MILES, Bruzual \& Charlot 2003 (BC03), Charlot \& Bruzual 2019 (CB19), and FSPS, we find systematic differences in mass-to-light ratios at $3.6~{\rm \mu m}$ ($M_{\ast}/L_{\rm 3.6\mu m}$) among the SPS models.
In particular, BC03 and CB19 yield mass-to-light ratios on average $\sim0.2-0.3~{\rm dex}$ lower than those from E-MILES and FSPS.
These mass-to-light ratios strongly correlate with stellar age, indicating a significant impact of young stellar populations on stellar mass measurements.
Our analysis, incorporating fiducial masses and simulated SPHEREx data, identifies the $1.6~{\rm \mu m}$ band as the optimal wavelength for stellar mass estimation, with the lowest scatter ($0.15-0.20~{\rm dex}$) of the stellar mass.
This scatter can be further reduced to $0.10-0.12~{\rm dex}$ across all SPS models by incorporating optical and SPHEREx colors.
These results can provide guidance for measuring the stellar masses of the numerous nearby galaxies that SPHEREx will survey.

\end{abstract}

\section{Introduction}

Stellar masses of galaxies ($M_{\ast}$) serve as a valuable proxy for understanding their evolutionary history, as they reflect the cumulative result of star formation and the gravitational distribution of stellar components.
Stellar masses are closely linked to other stellar and dynamical properties of galaxies.
For instance, \citet{kau03} demonstrated that stellar masses can help distinguish recent starburst galaxies (with lower mass and stellar concentrations) and quiescent galaxies (with higher mass and stellar concentrations), with a dividing boundary at ${\rm log}~M_{\ast}/M_{\odot}\sim10.5$ \citep[][]{pen10, van14}.
Furthermore, \citet{tre04} found a strong positive correlation between stellar masses and gas-phase metallicity, indicative of the ``mass--metallicity relation" \citep{kew08, zah14}.
Additionally, it has been reported that stellar masses of galaxies have close correlations with their halo and black hole masses, suggesting that these relations are connected to galaxy growth \citep{kor13, wec18, beh19, gre20, seo20}.
These relationships indicate that the stellar masses of galaxies are strongly dependent on their star formation history (SFH), internal structures, chemical compositions, halo properties, and black hole masses, making stellar mass a fundamental parameter for understanding the distributions of galaxy properties.
Therefore, it is essential to examine stellar masses and related properties, such as the star formation rate density, local metallicity, and stellar ages \citep[][and references therein]{san20}, on spatially resolved scales to gain deeper insights into galaxy evolution.

In observational studies, various techniques have been developed to estimate stellar masses or mass-to-light ratios ($M_{\ast}/L_{\lambda}$) for large samples of galaxies.
A commonly used method is based on empirical relations between mass-to-light ratios and broad-band photometric data.
\citet{bel01} constructed a simple linear relation of ${\rm log}~M_{\ast}/L_{\lambda}=a_{\lambda}+b_{\lambda}\times{\rm color}$ for the Johnson/Cousins $BVRI$ and $JHK$ bands, which was later extended to the Sloan Digital Sky Survey (SDSS) bands by \citet{bel03}, as detailed in their Appendix A2.
These relations show that mass-to-light ratios at longer near-infrared (NIR) wavelengths have lower slopes concerning color than those in the optical bands.
Leveraging the low variations in mass-to-light ratios in the infrared (IR) regime, several studies have used IR photometric data from the Two Micron All Sky Survey (2MASS) $K_{s}$ or Spitzer/IRAC $3.6~{\rm \mu m}$ and $4.5~{\rm \mu m}$ to estimate stellar masses \citep{li07, zhu10, esk12}.
The Spitzer Survey of Stellar Structure in Galaxies \citep[$\rm{S^{4}G}$;][]{mei12, mei14, que15} produced stellar mass maps for 1,627 nearby galaxies using Spitzer $3.6~{\rm \mu m}$ and $4.5~{\rm \mu m}$ fluxes, which assumed a constant mass-to-light ratio of $M_{\ast}/L_{\rm 3.6~\mu m}=0.6$ with accounting for non-stellar contributions in the NIR bands.
Additionally, the Wide-Field Infrared Survey Explorer (WISE) all-sky IR data \citep{wri10} have been used to derive scaling relations for stellar masses based on WISE luminosities and colors \citep{hwa12, jar13, clu14, jar23}.
\citet{ler19} further introduced a bounded power-law model for mass-to-light ratios at WISE $3.4~{\rm \mu m}$ (W1) or Spitzer/IRAC $3.6~{\rm \mu m}$ that incorporates specific star formation rates and colors from the Galaxy Evolution Explorer and WISE for approximately 16,000 local galaxies (see their Appendices A4 and A5).

Alternative methods for estimating stellar masses involve spectral energy distribution (SED) modeling based on the selection of stellar population synthesis (SPS) models (as discussed in {\color{blue} \textbf{Section \ref{sec:sps}}}).
These methods offer a more fundamental approach than empirical relations.
Broad-band SED fitting has proven to be effective for deriving stellar population properties, such as stellar masses, ages, and metallicities, for galaxies from the nearby to the high-redshift universe \citep{sal07, muz09, tay11, mou13, sal16, Lee24, wan24}.
Notably, as part of the Galaxy And Mass Assembly project, \citet{bel20} provided fiducial stellar masses, which have been used for developing recent WISE scaling relations \citep{jar23}, utilizing multiwavelength photometric data ranging from ultraviolet (UV) to far-infrared (FIR) wavelengths.
High-quality spectra also allow for the estimation of stellar population parameters through full-spectrum fitting, which can effectively recover non-parameterized SFHs of galaxies by combining spectral templates from individual simple stellar populations (SSPs) included in specific SPS models.
This method can alleviate potential systematic uncertainties associated with the parameterized SFH model selection, which is mostly adopted in broad-band SED fitting \citep{wal11, sim14, low20}.
Full-spectrum fitting techniques have been successfully used to estimate stellar population parameters for spectroscopically observed galaxies \citep{cid05, ocv06, kol09, ko16, wil17, cap23}.
Although these techniques also exhibit systematic differences depending on the choice of underlying SPS models \citep{gon10, ge19}, they are highly effective for robust stellar mass estimation because they are less sensitive to the systematic degeneracy of other parameters, such as metallicity, dust extinction, and SFH models.

Each stellar mass estimation method has its strengths and limitations, but this study employs spectrum-derived methods tailored to the objectives of the upcoming Spectro-Photometer for the History of the Universe, Epoch of Reionization, and Ices Explorer (SPHEREx).
In the near future, SPHEREx will provide all-sky IR spectrophotometric data covering wavelengths from $0.75~{\rm \mu m}$ to $5~{\rm \mu m}$.
One of its scientific objectives is to map the distribution of stellar population properties in nearby galaxies, including their stellar masses and specific star formation rates \citep{dor16, dor18}. 
SPHEREx is expected to provide resolved stellar mass maps which are more informative than integrated stellar masses, as it can account for spatial variations in dust extinction and star formation activity, particularly in late-type galaxies \citep{zib09, pac19a, pac19b}.
These resolved stellar mass maps facilitate the detailed study of the formation and evolution of galactic structures such as bulges, disks, bars, and spiral arms \citep{gad09, sal15}.
Specifically, they can serve as an important basis for investigating 
the radial mass distribution of disks \citep{fre70, poh06, erw08}, 
stellar migration induced by bars and spiral arms \citep{sel02, ros08, min10}, 
the detailed structures of bars and spiral arms by constructing gravitational potential maps \citep{but01, lau02}, 
and dark matter distribution in galaxies \citep{deB08, oh11, tam12}.

The NIR wavelength range of SPHEREx ($0.75-5~{\rm \mu m}$) is well-suited for investigating stellar mass distributions in galaxies for the following reasons.
First, the light within this wavelength range is dominated by the continuum emission from old stars, which are the main contributors to the stellar mass of galaxies.
Second, IR light is less affected by interstellar extinction compared with optical light, providing a clearer view of stellar emissions.
Third, the stellar mass-to-light ratios in the IR regime are relatively insensitive to variations in stellar ages and metallicities \citep{mei14, nor14}.
Given these advantages, SPHEREx will offer an excellent opportunity for generating precise resolved stellar mass maps for a large number of nearby galaxies.
Therefore, it is necessary to develop a preemptive strategy for stellar mass estimation tailored to the capabilities of SPHEREx.

Considering these aspects, this paper presents a pilot study on resolved stellar mass estimation of nearby late-type galaxies in preparation for the upcoming SPHEREx mission.
Late-type galaxies serve as useful laboratories for studying the relationships between resolved stellar masses and stellar populations because their mass-to-light ratios in the IR regime exhibit greater spatial variation compared with those in early-type galaxies \citep{ler19, jar23}.
The primary objectives of this work can be summarized as follows:
First, we generate resolved stellar mass maps of nearby galaxies with the SPHEREx pixel size of $6\farcs2$, utilizing full-spectrum fitting with Multi Unit Spectroscopic Explorer (MUSE) integral field spectroscopic data, derived as part of the Physics at High Angular Resolution in Nearby Galaxies (PHANGS) survey \citep[PHANGS-MUSE;][]{ems22}.
We use these resolved stellar mass maps to examine the dependence of mass-to-light ratios on stellar population properties (e.g., ages and metallicities).
Second, we investigate systematic differences in stellar mass estimations among different SPS models and compare the predicted mass-to-light ratios in IR bands across these SPS models.
Third, we propose new stellar mass estimators based on the NIR continuum and optical colors, which will facilitate the estimation of resolved stellar masses in nearby galaxies using future SPHEREx data.

This paper is structured as follows.
{\color{blue} \textbf{Section \ref{sec:data}}} describes the archival data used to create resolved stellar mass maps for nearby galaxies and compare them with previous results, including data from the PHANGS-MUSE Integral Field Unit (IFU), the ${\rm S^{4}G}$ survey, and WISE images.
{\color{blue} \textbf{Section \ref{sec:sps}}} provides a detailed explanation of the four SPS models used in this study.
{\color{blue} \textbf{Section \ref{sec:analysis}}} outlines our methodology for full-spectrum fitting to estimate resolved stellar masses and tests the influence of SPHEREx-like binning on the full-spectrum fitting results.
Subsequently, {\color{blue} \textbf{Section \ref{sec:model}}} present the distributions of stellar population parameters, such as ages, metallicities, mass-to-light ratios, and IR colors depending on the SPS models.
In {\color{blue} \textbf{Section \ref{sec:spherex}}}, we develop new IR scaling relations, including SPHEREx luminosity and color, for accurate stellar mass prediction.
{\color{blue} \textbf{Section \ref{sec:summary}}} summarizes our main conclusions.
We also compare the resolved stellar masses obtained in this study with previous results in {\color{blue} \textbf{Appendix A}}.
Throughout the paper, we use the AB magnitude system for the SDSS and SPHEREx bands, and the Vega magnitude system for the 2MASS, IRAC, and WISE bands.

\section{The PHANGS-MUSE Survey Data}
\label{sec:data}


We utilized publicly available optical IFU data obtained with MUSE on the Very Large Telescope (VLT) from the PHANGS-MUSE survey\footnote[1]{\url{https://www.canfar.net/storage/vault/list/phangs/RELEASES/PHANGS-MUSE}}.
The data cover a wavelength range of $4750-9350~{\rm \AA}$, with a spectral resolution ranging from $\lambda/\Delta\lambda_{\rm FWHM}\sim1600$ at the blue end to $\lambda/\Delta\lambda_{\rm FWHM}\sim3600$ at the red end.
The PHANGS-MUSE sample includes 19 star-forming disk galaxies in the local universe within $20~{\rm Mpc}$ \citep{ems22}.
Multiple VLT/MUSE pointings using the wide-field mode ($1\arcmin \times 1\arcmin$) were used to cover most of the stellar disk regions of these galaxies, including galactic centers, spiral arms, interarm regions, and bars.
The released IFU datacubes were calibrated for astrometry and photometry and coadded with a homogenized spatial resolution, resulting in full width at half maximums (FWHMs) of $0\farcs6-1\farcs2$ for all pointings of each galaxy.
This sample is well-suited for investigating spatial variations in stellar populations \citep{pes23} and star formation activity \citep{lom24} in nearby galaxies.

For this study, we used the released datacubes of all 19 PHANGS-MUSE galaxies to create resolved stellar mass maps with SPHEREx pixel sampling.
To match the spatial resolution of the SPHEREx data, we 
resampled the spaxels from their original resolution of $0\farcs2~{\rm pixel^{-1}}$ to the SPHEREx pixel size of $6\farcs2~{\rm pixel^{-1}}$ by summing the pixel values.
Given that the SPHEREx pixel size is substantially larger than that of the Voronoi tessellations (with a target signal-to-noise ratio of 35) used in the PHANGS-MUSE released data products, the rebinned spaxels exhibit high median signal-to-noise ratios at $5000-5500~{\rm \AA}$, ranging from $\sim50$ in the galactic outskirts to $\sim2000$ in the central regions.
This re-binning process generates spatially resolved optical spectra with high spectral resolutions, complementing the SPHEREx dataset for nearby galaxies and providing valuable data for developing optimal strategies for resolved stellar mass estimation with SPHEREx.

In this study, we corrected for foreground extinction using the \texttt{extinction} package\footnote[2]{\url{https://github.com/kbarbary/extinction}} \citep{bar16}, applying the dust extinction law from \citet{car89} and the color excess magnitudes of the Milky Way ($E(B-V)_{\rm MW}$) from \citet{sch11}.

\begin{figure*}
\centering
\includegraphics[width=1.0\textwidth]{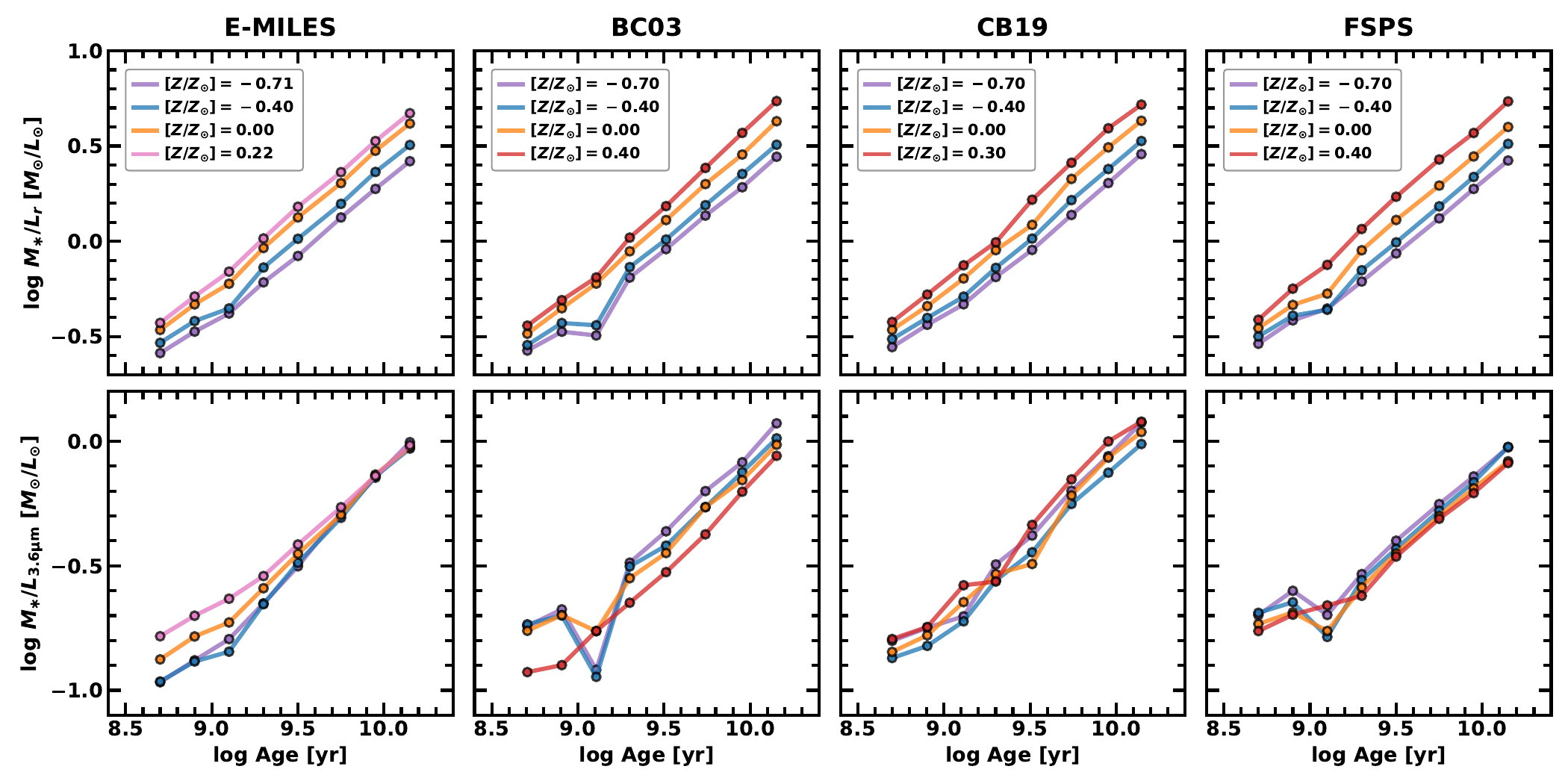}
\caption{
Stellar mass-to-light ratios as a function of age, derived from the spectral templates from the four SPS models: E-MILES (first column), BC03 (second column), CB19 (third column), and FSPS (fourth column).
The upper set of four panels displays the mass-to-light ratios calculated in the SDSS $r$ band, while the lower set shows those obtained in Spitzer/IRAC channel 1 ($3.6~{\rm \mu m}$).
Each panel is color-coded according to the metallicity bin from the SPS models.
\label{fig:temp}}
\end{figure*}

\section{SPS Models}
\label{sec:sps}

SPS models are constructed from SSPs, which include fundamental components such as the initial mass function (IMF), stellar isochrones, and spectral templates of stellar light \citep{con13}.
These SPS models provide theoretical predictions of SEDs for unresolved stellar light, accounting for variations in stellar age, metallicity, dust absorption, or SFH.
In observational studies, SPS models are crucial for investigating the properties of stellar populations in star clusters or galaxies.
Therefore, several well-established SPS models have been widely used for SED fitting with photometric data or full spectral continuum fitting for various stellar systems.

In this study, we selected four SPS models that cover the IR wavelengths: E-MILES \citep{vaz16}, 
BC03 \citep{bru03}, 
Charlot \& Bruzual 2019 \citep[CB19;][]{pla19}, 
and FSPS \citep{con09, con10}.
Our objective herein was to explore the dependence of SPS models on stellar mass estimation, focusing on mass-to-light ratios in the IR wavelengths relevant to SPHEREx.
The SPS models considered in this study consist of spectral templates corresponding to specific ages and metallicities, each based on a single-burst SFH (i.e., simple stellar populations; SSPs).
For all SPS models, we adopted the Chabrier IMF \citep{cha03}, with a mass range from $0.1~M_{\odot}$ to $100~M_{\odot}$, and used air wavelengths.

Another useful IR-covering model, Maraston 2005 \citep[M05;][]{mar05}, can be reasonably compared with BC03 due to its significantly higher NIR flux levels, which is attributed to different prescriptions for the thermally pulsating asymptotic giant branch (TP-AGB) phase stars.
However, we excluded M05 from our analysis because it provides stellar templates with lower spectral resolution and offers a limited choice of IMFs compared with the other SPS models, which could introduce systematic biases in stellar mass estimation.

To illustrate the differences among these SPS models, {\color{blue} \textbf{Figure \ref{fig:temp}}} shows the stellar mass-to-light ratios in the SDSS $r$ band and the Spitzer/IRAC ${\rm 3.6~\mu m}$ channel as functions of age and metallicity for SSPs with older stellar ages than ${\rm log~Age~(yr)>8.5}$ in the four SPS models.
Across all bands and SPS models, the mass-to-light ratios exhibit a strong positive correlation with stellar age.
In the $r$ band, mass-to-light ratios show a notable dependence on metallicity, with higher metallicity bins yielding higher mass-to-light ratios.
The variation in $M_{\ast}/L_{r}$ between the highest and lowest metallicity bins is $0.2-0.3~{\rm dex}$ across all SPS models, and these trends are consistent for a given age and metallicity.
In contrast, the mass-to-light ratios at ${\rm 3.6~\mu m}$ reveal systematic differences among the SPS models.
In particular, there are substantial variations in mass-to-light ratios at the intermediate age around 1 Gyr due to the different treatments of TP-AGB stars.
Even for older stellar populations (${\rm log~Age~(yr)}>9.5$), the metallicity dependence of mass-to-light ratios varies between models.
For instance, E-MILES shows minimal variation of mass-to-light ratios with metallicity, while other SPS models show systematic differences of $0.1–0.2~{\rm dex}$.
Interestingly, BC03 and FSPS show higher mass-to-light ratios at lower metallicities, consistent with trends observed in Figure 2 of \citet{mei14}, whereas CB19 shows the opposite metallicity dependence.
These findings suggest that the choice of SPS models can introduce larger uncertainties in predicting mass-to-light ratios at NIR wavelengths compared to optical wavelengths.
These discrepancies may lead to systematic differences in the stellar mass estimates of galaxies, particularly when using SPHEREx, underscoring the importance of this study.
Detailed descriptions of the four SPS models are provided in the following subsections.

\subsection{E-MILES with Superyoung Models}
\label{sec:emiles}

For the spectral analysis in this study, we obtained the E-MILES templates from the MILES cloud storage\footnote[3]{\url{https://cloud.iac.es/index.php/s/aYECNyEQfqgYwt4}}.
The E-MILES model provides stellar spectral templates covering wavelengths ranging from $0.16~{\rm \mu m}$ to $5~{\rm \mu m}$, which is an extended version of the MILES stellar library \citep{san06}.
These templates offer a high spectral resolution of $\Delta\lambda_{\rm FWHM}\sim2.5~{\rm \AA}$ in the optical wavelength range up to $0.9~{\rm \mu m}$, but the resolution decreases to $\Delta\lambda_{\rm FWHM}\sim 15-20~{\rm \AA}$ in the Spitzer/IRAC $3.6~{\rm \mu m}$ and $4.5~{\rm \mu m}$ bands \citep[see Figure 8 in][]{vaz16}.
We selected the E-MILES templates based on the Padova 2000 isochrones \citep{gir00} and a scaled-solar model for metallicity, assuming ${\rm [Fe/H]=[Z/Z_{\odot}]}$.
For our stellar population analysis, we utilized spectral templates for 12 age bins of [0.0708, 0.1122, 0.1778, 0.2818, 0.5012, 0.7943, 1.2589, 1.9953, 3.1623, 5.6234, 8.9125, 14.1254] in Gyr and four metallicity bins of ${\rm [Z/Z_{\odot}]}=[-0.71, -0.40, 0.00, 0.22]$.

However, the E-MILES models lack spectral templates for stellar ages younger than $63~{\rm Myr}$, which can introduce systematic uncertainties when analyzing young stellar populations, as discussed in Section 6.3.2 of \citet{ems22}.
To address this limitation, we supplemented our analysis with ``superyoung'' models, as done in the previous PHANGS-MUSE study on stellar population properties by \citet{pes23}.
These superyoung E-MILES models differ from their older versions in the sense that they employ the Padova 1994 isochrones \citep{gir96} and provide a higher metallicity set with ${\rm [Z/Z_{\odot}]=0.41}$ instead of ${\rm [Z/Z_{\odot}]=0.22}$.
In addition, these templates do not include IR spectra, covering only wavelengths up to $0.9~{\rm \mu m}$.
This limitation could potentially affect the accuracy of mass-to-light ratios at IR wavelengths for regions with stellar ages younger than $10^{8}~{\rm yr}$.
Following \citet{pes23}, we included the additional superyoung models with five age bins of [0.0063, 0.0100, 0.0158, 0.0251, 0.0398] in Gyr and four metallicity bins of ${\rm [Z/Z_{\odot}]}=[-0.71, -0.40, 0.00, 0.41]$.

\subsection{BC03 Models}
\label{sec:bc03}

BC03 is one of the most widely used SPS models for predicting the distribution of stellar populations.
In this study, we utilized the updated 2016 version of the BC03 templates, computed using the GALAXEV code\footnote[4]{\url{https://www.bruzual.org/bc03}}.
The BC03 models cover a wide wavelength range from $91~{\rm \AA}$ to $160~{\rm \mu m}$.
They combine a high-resolution spectral library ($\Delta\lambda_{\rm FWHM}\sim3~{\rm \AA}$) from STELIB \citep{leb03} for the $3200-9500~{\rm \AA}$ range, alongside lower-resolution libraries ($\Delta\lambda_{\rm FWHM}\sim20~{\rm \AA}$) from BaSeL 3.1 \citep{wes02} and Pickles \citep{pic98} for the remaining wavelength range.
These models are based on the Padova 1994 isochrones, rather than the Padova 2000 isochrones, as detailed in \citet{bru03}.
BC03 provides spectral templates for a wide range of stellar ages from $0.1~{\rm Myr}$ to $20~{\rm Gyr}$.
For this study, we adopted templates for 17 age bins of [0.0063, 0.01, 0.0158, 0.0251, 0.04, 0.0719, 0.1139, 0.1805, 0.2861, 0.5088, 0.8064, 1.2781, 2.0, 3.25, 5.5, 9.0, 14.25] in Gyr and four metallicity bins of ${\rm [Z/Z_{\odot}]}=[-0.70, -0.40, 0.00, 0.40]$.

\subsection{CB19 Models}
\label{sec:cb19}

CB19\footnote[5]{\url{https://www.bruzual.org/CB19/CB19_chabrier/Mu100}} is the latest major revision of BC03, incorporating the PARSEC stellar evolutionary tracks \citep{bre12} instead of the older Padova isochrones.
These updated PARSEC tracks incorporate newly computed models for the evolutionary physics of TP-AGB stars \citep{mar13} and stellar winds from hot massive stars, such as OB-type and Wolf--Rayet stars \citep{che15}.
The CB19 models provide a broad wavelength range from $5.6~{\rm \AA}$ to $36,000~{\rm \mu m}$ by integrating various spectral libraries: MILES for wavelengths of $\lambda\sim3540-7350~{\rm \AA}$, STELIB for wavelengths of $\lambda=7351-8750~{\rm \AA}$, and BaSeL 3.1 combined with several TP-AGB models for wavelengths longer than $8750~{\rm \AA}$ \citep[see Appendix A in][and references therein]{san22}.
In this study, we used the CB19 templates including 17 age bins of [0.0063, 0.01, 0.016, 0.025, 0.040, 0.070, 0.110, 0.180, 0.275, 0.5, 0.8, 1.3, 2.0, 3.25, 5.5, 9.0, 14.0] in Gyr and four metallicity bins of ${\rm [Z/Z_{\odot}]}=[-0.70, -0.40, 0.00, 0.30]$.

\subsection{FSPS Models}
\label{sec:fsps}

The FSPS models, developed by \citet{con09}, offer great flexibility in terms of IMFs, isochrones, spectral libraries, and physical treatments for stellar phenomena such as the horizontal branch (HB), blue stragglers, and TP-AGB stars.
For this study, we compiled \textsc{FSPS v3.2}\footnote[6]{\url{https://github.com/cconroy20/fsps}} and generated the default FSPS models using the \texttt{python-fsps}\footnote[7]{\url{https://github.com/dfm/python-fsps}} package.
These templates are based on the spectral libraries of MILES ($\lambda\sim3600-7400~{\rm \AA}$) and BaSeL 3.1 ($\lambda=91~{\rm \AA}-160~{\rm \mu m}$, excluding the MILES coverage), the MIST isochrones \citep{cho16}, and the interstellar dust emission model from \citet{dra07}.
We used the FSPS models for 14 age bins of [0.0063, 0.01, 0.0158, 0.0251, 0.0398, 0.0708, 0.1122, 0.1778, 0.2818, 0.5012, 0.7943, 1.2589, 1.9953, 3.1623, 5.6234, 8.9125, 14.1254] in Gyr and four metallicity bins of ${\rm [Z/Z_{\odot}]}=[-0.70, -0.40, 0.00, 0.40]$.

\section{Analysis}
\label{sec:analysis}

\subsection{Full-Spectrum Fitting}
\label{sec:ppxf}

Full-spectrum fitting is a powerful technique for quantifying stellar population parameters, such as age, metallicity, and mass, and for reconstructing the detailed SFH of stellar systems.
In this study, we employed the Penalized PiXel-Fitting \citep[pPXF;][]{cap04, cap17, cap23} program\footnote[8]{\url{https://pypi.org/project/ppxf/9.1.1}} for full-spectrum fitting, which is known for its superior performance in recovering stellar population parameters and computational efficiency \citep{ge18, woo24}.
For our implementation of pPXF, we applied the four different SPS models described in {\color{blue} \textbf{Section \ref{sec:sps}}} to the optical IFU datacubes rebinned to the SPHEREx pixel size, as described in {\color{blue} \textbf{Section \ref{sec:data}}}.

Prior to running pPXF, we manually masked all emission lines, including Balmer lines and strong forbidden line doublets (e.g., [\ion{O}{3}]$\lambda\lambda4959,5007$, [\ion{N}{2}]$\lambda\lambda6548,6584$, and [\ion{S}{2}]$\lambda\lambda6717,6731$), to minimize contamination from nebular emission in the stellar light components.
Following the methods used in the previous PHANGS-MUSE works \citep{ems22, pes23}, we implemented the pPXF tasks in two steps, i.e., deriving the stellar kinematics and estimating the stellar populations with the fixed kinematic parameters.
For both steps, we used a spectral range of $4850-7000~{\rm \AA}$ to avoid unstable sky subtraction at longer wavelengths, as noted by \citet{ems22}.
These procedures ensured consistency with previous PHANGS-MUSE results.

\begin{figure*}
\centering
\includegraphics[width=0.82\textwidth]{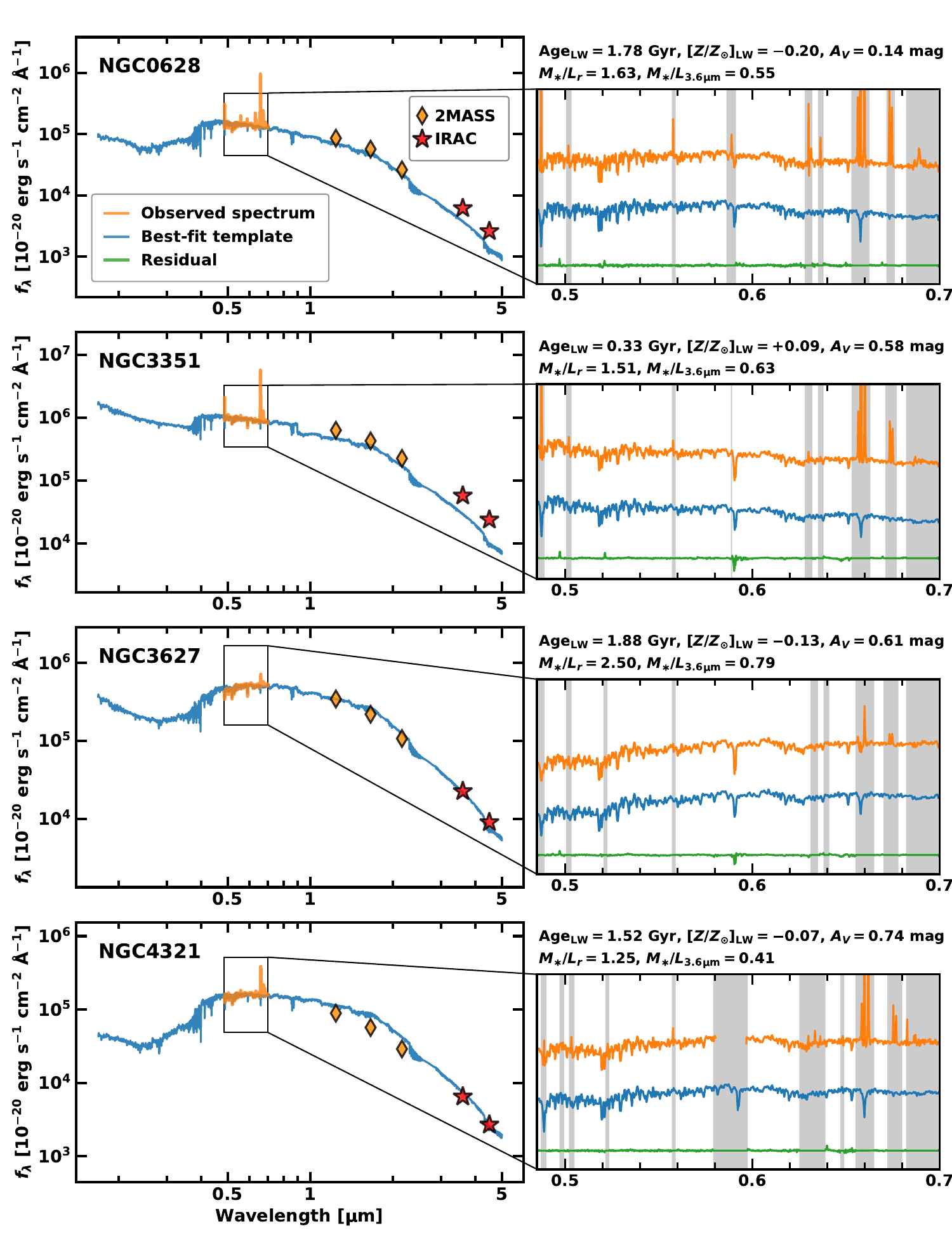}
\caption{
Example SEDs of rebinned spaxels in the central regions of the four PHANGS-MUSE galaxies.
The left panels display the observed spectra (orange) and best-fit templates from pPXF (blue) over the full wavelength range of the SPS model ($\lambda\sim0.2-5.0~{\rm \mu m}$), along with the photometric data points from 2MASS $JHK_{s}$ (orange diamonds), and Spitzer/IRAC $3.6$, $4.5~{\rm \mu m}$ (red star symbols).
The zoomed-in panels on the right show the SEDs within the pPXF fitting range of $\lambda=0.485-0.7~{\rm \mu m}$.
The observed spectra and best-fit templates are depicted in the left panels, with the residual SEDs represented by green curves.
Gray-shaded regions indicate the masked wavelength ranges.
The stellar population parameters derived from the best-fit pPXF results are displayed at the top of each zoomed-in panel.
These results were obtained using the E-MILES models.
\label{fig:spec}}
\end{figure*}

To examine the stellar kinematics, we measured four parameters: the radial velocity along the line of sight ($v_{\rm rad}$), stellar velocity dispersion ($\sigma_{v}$), and third- and fourth-order Gauss--Hermite coefficients ($h_{3}$ and $h_{4}$, respectively).

The kinematics was determined through three iterations of the pPXF process to obtain stable solutions.
The first iteration provided initial guesses for the kinematics and constructed noise models by comparing the best-fit templates with the observed data.
The second iteration refined these initial guesses and noise models, and the third iteration determined the final kinematic parameters.
Throughout these processes, we used additive Legendre polynomials with a degree of 12 (\texttt{degree=12}), no multiplicative Legendre polynomials (\texttt{mdegree=0}), and a regularization error of 0.2 (\texttt{regul=5})
, consistent with earlier PHANGS-MUSE works.
Following \citet{ems22}, we used only the E-MILES library with the BaSTI isochrones \citep{pie04} to explore the kinematics, which differs from the template sets used for stellar population analysis.

To investigate the stellar populations, we applied the four different SPS models in the pPXF tasks, using the fixed kinematic parameters derived from the previous step.
Similar to the kinematic analysis, 
we conducted four iterations of the pPXF tasks.
The first and second iterations were focused on generating and improving the noise models for the input spectra.
The third iteration was aimed at deriving internal extinction magnitudes ($A_{V}$) using the \texttt{reddening} keyword, applying the \citet{cal00} extinction law.
The spectra were then corrected for internal extinction using the derived $A_{V}$ values.
In the final iteration, we used the extinction-corrected spectra to derive the weighting coefficients of the input spectral templates, which were then used to determine the stellar population parameters.
Throughout these processes, we used no additive polynomials (\texttt{degree=-1}), multiplicative polynomials with a degree of 12 (\texttt{mdegree=12}), and no regularization (\texttt{regul=0}), except during the third iteration for internal extinction.
These configurations were consistently set according to \citet{pes23}.
In the third iteration, we used no multiplicative polynomials (\texttt{mdegree=0}) and activated the \texttt{reddening} option \citep[similar to Choice-1 in][]{lee24}.

{\color{blue} \textbf{Figure \ref{fig:spec}}} illustrates example results of the full-spectrum fitting process for several rebinned spaxels in the central regions of the PHANGS-MUSE galaxies.
The zoomed-in panels show the best-fit model spectra (blue lines) compared with the observed spectra (orange lines) within the given wavelength ranges.
As all SPS models used in this study cover NIR wavelengths up to $5~{\rm \mu m}$, the best-fit model spectra can be extended to these wavelengths by applying the output stellar population parameters and the weighting coefficients of the input spectral templates.
The zoomed-out panels display the best-fit results across the full wavelength ranges of the SPS templates, from UV to NIR.
For these panels, the photometric data of 2MASS and Spitzer/IRAC were obtained from the pixel values in the images reprojected to the SPHEREx pixel size.
These wavelength extensions of the best-fit results were used to estimate the resolved stellar mass and mass-to-light ratios at NIR wavelengths, as will be mentioned in {\color{blue} \textbf{Section \ref{sec:Mstar}}}.
For our analysis, we excluded the SPHEREx-binned spaxels with a signal-to-noise ratio below 155 (equivalent to a mean signal-to-noise ratio of 5 for the original unbinned pixel scale) and unreliable kinematics measurements with $v_{\rm rad}<500~{\rm km~s^{-1}}$ or $\sigma_{v}>200~{\rm km~s^{-1}}$.

\begin{figure*}
\centering
\includegraphics[width=1.0\textwidth]{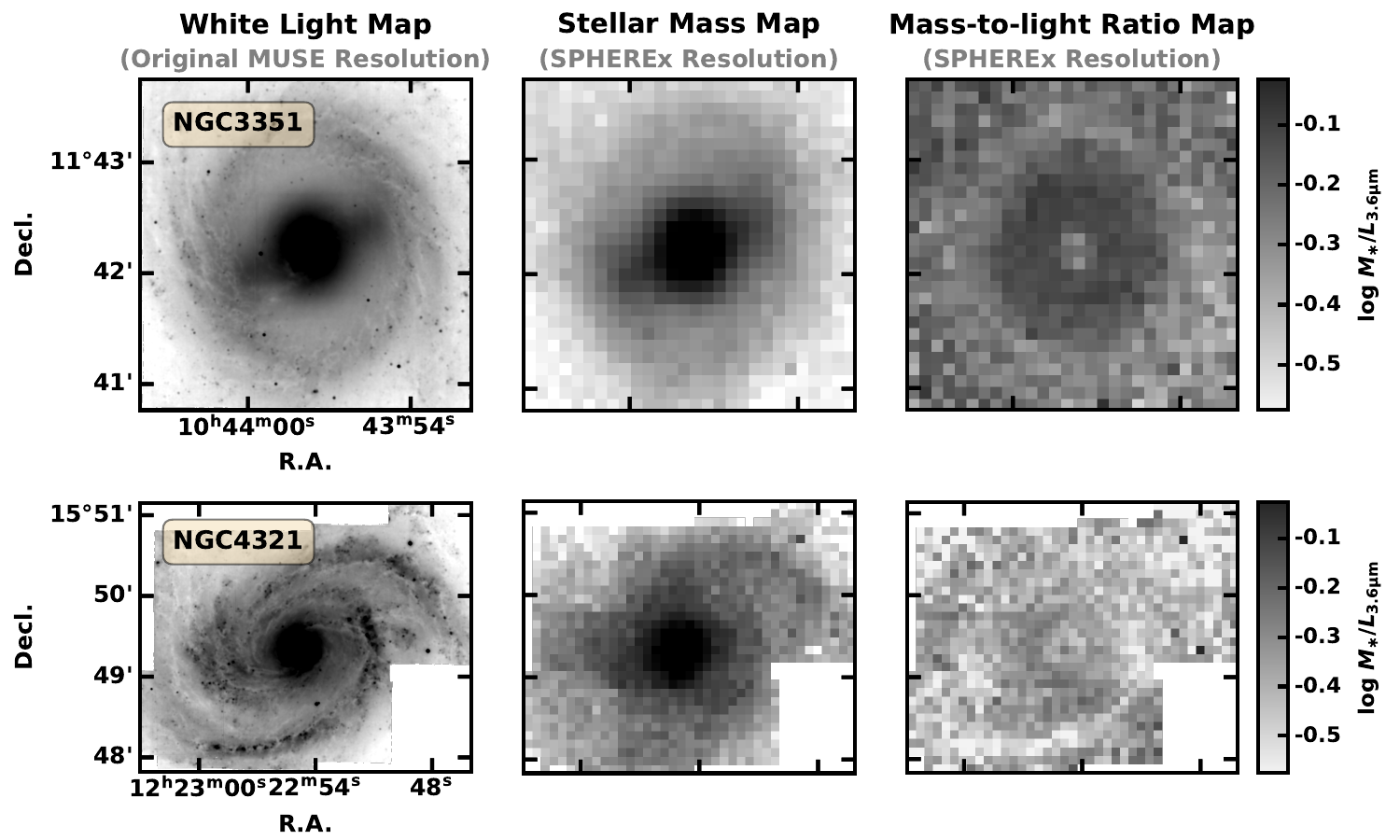}
\caption{
Example white light maps (left column), resolved stellar mass maps (middle column), and $3.6~{\rm \mu m}$ mass-to-light ratio maps (right column) for two PHANGS-MUSE galaxies.
The white light maps are generated from the original PHANGS-MUSE datacubes by summing the flux across all wavelengths, providing a comparison with the SPHEREx-binned maps.
The resolved stellar mass maps and mass-to-light ratio maps are based on the SPHEREx-binned images, which were used to determine the fiducial stellar masses in this study.
\label{fig:maps}}
\end{figure*}

\subsection{Estimation of Resolved Stellar Masses and Mass-to-Light Ratios at NIR}
\label{sec:Mstar}

The pPXF routines provide luminosity-weighted coefficients ($w_{i}^{L}$) for the $i$-th SSP template in the input SPS models for all spaxels of the SPHEREx-rebinned PHANGS-MUSE datacubes.
To calculate stellar mass-to-light ratios from the best-fit pPXF results, we converted the output luminosity weights to mass weights ($w_{i}^{M}$) through division by the mean flux of each SSP template within a normalizing wavelength range of $\lambda=5070-5950~{\rm \AA}$, corresponding to the typical FWHM range of the $V$-band.
Subsequently, we derived the mass-to-light ratio for a specific band ($M_{\ast}/L_{\lambda}$) from the best-fit pPXF results of each spaxel using the following formula:
\begin{equation}
    \frac{M_{\ast}}{L_{\lambda}}=\frac{\Sigma_{i} (w_{i}^{M}\times m_{\ast, i})}{\Sigma_{i} (w_{i}^{L}\times L_{\lambda, i})}
    \label{eqn:lw_to_mw}
\end{equation}
where $m_{\ast,i}$ represents the stellar mass of the $i$-th SSP spectral template in solar mass ($M_{\odot}$), and $L_{\lambda,i}$ is the specific luminosity at a given band for the $i$-th SSP template in specific solar luminosity ($L_{\lambda,\odot}$).
The values for $m_{\ast,i}$ are predetermined for all SSP templates by each SPS model, including the masses of living stars and stellar remnants such as white dwarfs, neutron stars, and black holes.
The luminosity at a specific band for the best-fit template was computed using the \texttt{PyPhot} package.

Using {\color{blue} \textbf{Equation \ref{eqn:lw_to_mw}}}, we computed the stellar mass-to-light ratios for all spaxels, ranging from the SDSS optical bands to the Spitzer/IRAC NIR bands.
The stellar mass of each spaxel was determined by multiplying the specific luminosity in the SDSS $r$ band by the corresponding mass-to-light ratio derived from its best-fit SED template.
Through these procedures, we created resolved stellar mass maps for the 19 PHANGS-MUSE galaxies, with an example shown in {\color{blue} \textbf{Figure \ref{fig:maps}}}.
These maps serve as the fiducial stellar masses used throughout this study.
We did not account for the inclination effects of disk galaxies in this study.

\begin{figure*}
\centering
\includegraphics[width=0.8\textwidth]{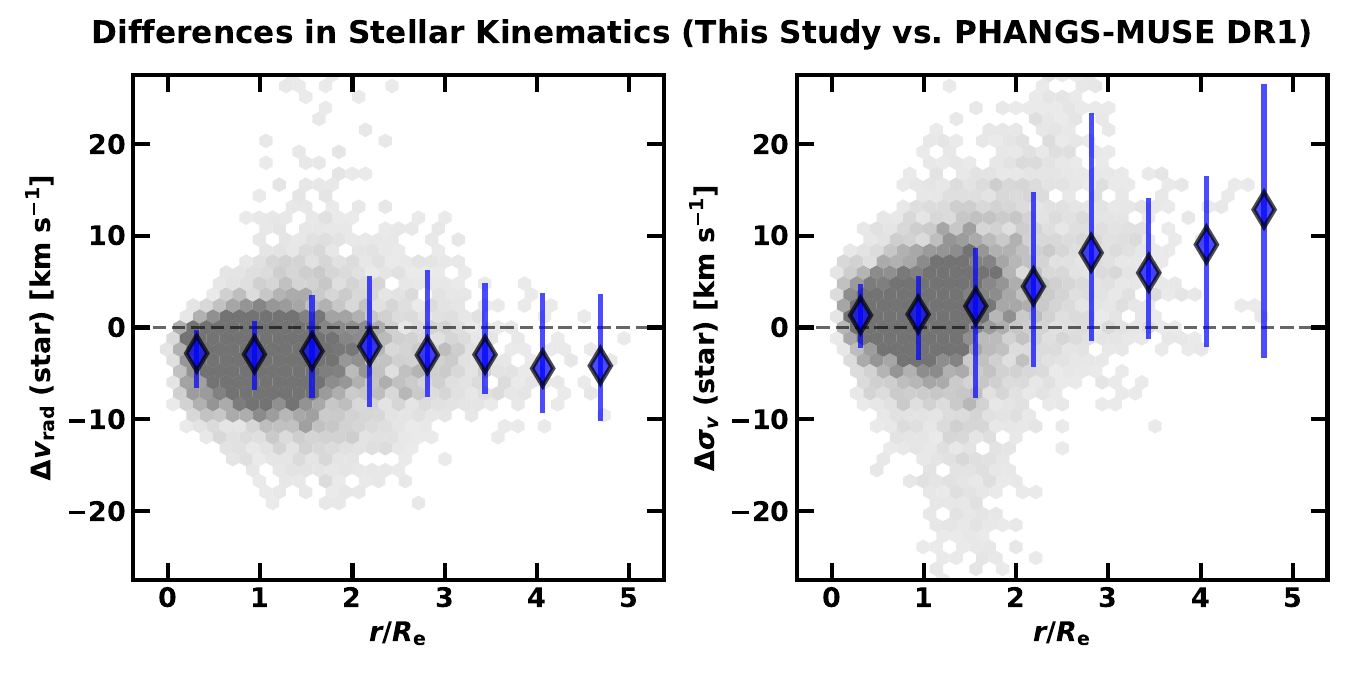}
\caption{
Comparison of stellar kinematics between this study and a previous PHANGS-MUSE study \citep{ems22}, both based on the E-MILES model.
The two panels show the differences in radial velocities (left) and velocity dispersions (right), plotted as a function of galactocentric radius normalized by the effective radius of each galaxy.
These differences are calculated by subtracting the luminosity-weighted mean kinematic parameters of Voronoi-binned spaxels (PHANGS-MUSE data release 1) from the values of SPHEREx-rebinned spaxels (this study).
Blue diamonds with error bars represent the median differences within the radial bins, alongside the 16th and 84th percentiles of their distributions.
\label{fig:comp_kin}}
\end{figure*}

\subsection{Tests of the Binning Effects}
\label{sec:binning}

Since the full-spectrum fitting in this study was performed on SPHEREx-binned IFU data, it is important to assess the potential influence of this binning on the pPXF results.
To quantify the effect of SPHEREx binning, we compared the stellar kinematics and stellar population parameters obtained in this study with those from previous PHANGS-MUSE analyses \citep{ems22, pes23}, which were based on Voronoi-binned datacubes.
Although stellar kinematics cannot be measured with SPHEREx data due to its low spectral resolution, this kinematics comparison is necessary for testing the reliability of the stellar population parameters because stellar kinematics can affect the stellar population analysis during the pPXF fitting process.
For the kinematics comparison, we used the value-added maps from PHANGS-MUSE data release v1.0, which provide stellar radial velocity and velocity dispersion maps derived using the pPXF configurations specified by \citet{ems22}.
For stellar populations, we digitized the data of luminosity-weighted stellar ages and metallicities from Figures 2 and 7 of \citet{pes23}, as these data were not publicly available.
To maintain consistency with previous studies, all comparative analyses described in this section were based on the E-MILES models.

\begin{figure*}
\centering
\includegraphics[width=0.8\textwidth]{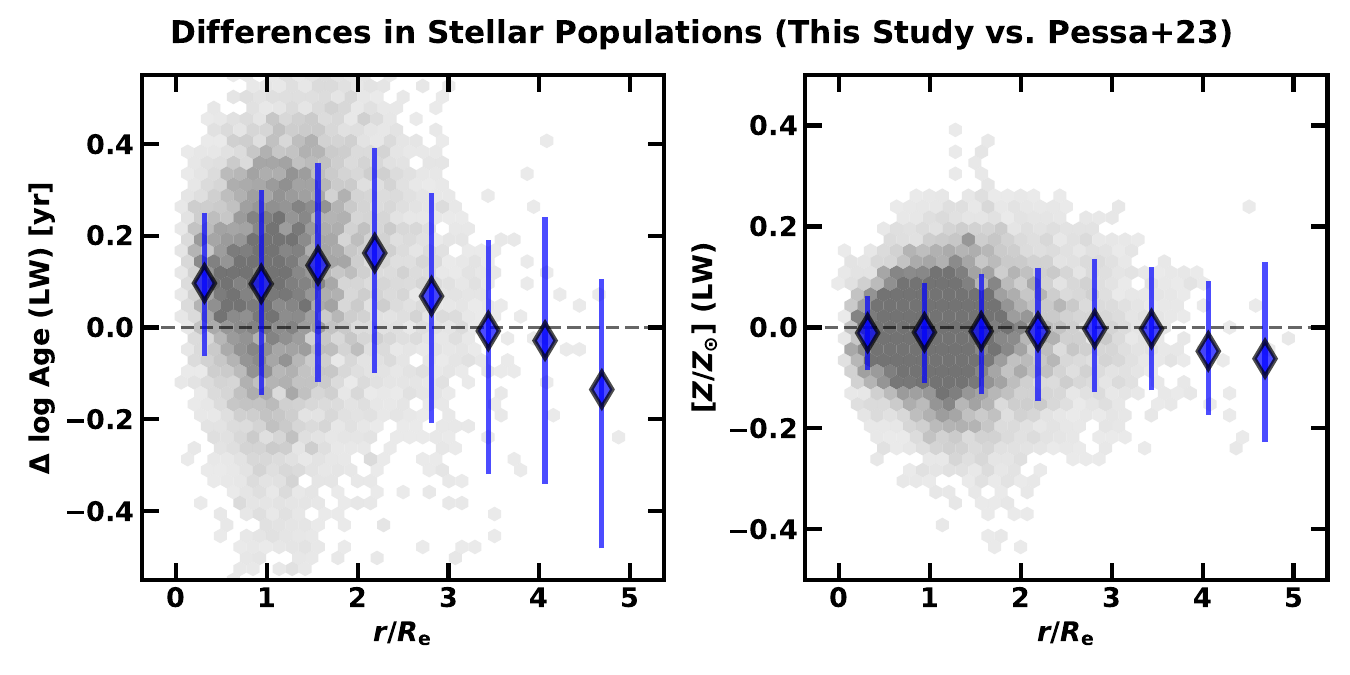}
\caption{
Comparison of stellar population parameters between this study and a previous PHANGS-MUSE study \citep{pes23}.
The two panels show the differences in stellar ages (left) and metallicities (right), with the same x-axis as in {\color{blue} \textbf{Figure \ref{fig:comp_kin}}}.
The data for ages and metallicities were digitized from Figures 2 and 7 in the work of \citet{pes23}, respectively.
The symbols have the same meaning as in {\color{blue} \textbf{Figure \ref{fig:comp_kin}}}.
\label{fig:comp_stp}}
\end{figure*}

{\color{blue} \textbf{Figure \ref{fig:comp_kin}}} presents comparisons of stellar radial velocity (left) and velocity dispersion (right) for 19 PHANGS-MUSE galaxies as a function of galactocentric distance, normalized by effective radius \citep[Table 4 in ][]{ler21}.
The radial velocities derived from the SPHEREx-binned data in this study exhibit a slight underestimation compared with the Voronoi-binned data across all radial bins, with a median discrepancy of $\sim -5~{\rm km~s^{-1}}$.
Given that the systematic radial velocities for the PHANGS-MUSE sample range from $650~{\rm km~s^{-1}}$ to $2400~{\rm km~s^{-1}}$, this discrepancy is considered negligible.
In terms of velocity dispersion, the results from this study show a slight overestimation of $\Delta\sigma_{v}\lesssim5~{\rm km~s^{-1}}$ within the regions up to $r/R_{\rm e}\lesssim2$, consistent with those from the previous study.
However, velocity dispersions from the SPHEREx-binned data tend to be higher in the outer regions than those from original PHANGS-MUSE data, with a discrepancy up to $\Delta\sigma_{v}\sim10~{\rm km~s^{-1}}$ at $r/R_{\rm e}>3$.
This overestimation may be associated with the smearing effect due to the SPHEREx-binning process.
This discrepancy is non-negligible, considering the median velocity dispersion of $49~{\rm km~s^{-1}}$ and its uncertainty of $17~{\rm km~s^{-1}}$ from the original datacubes.
Nonetheless, the effect of SPHEREx binning on velocity dispersion appears to be limited to a small number of pixels in the outer regions ($r/R_{\rm e}>3$).

\begin{figure*}
\centering
\includegraphics[width=1.0\textwidth]{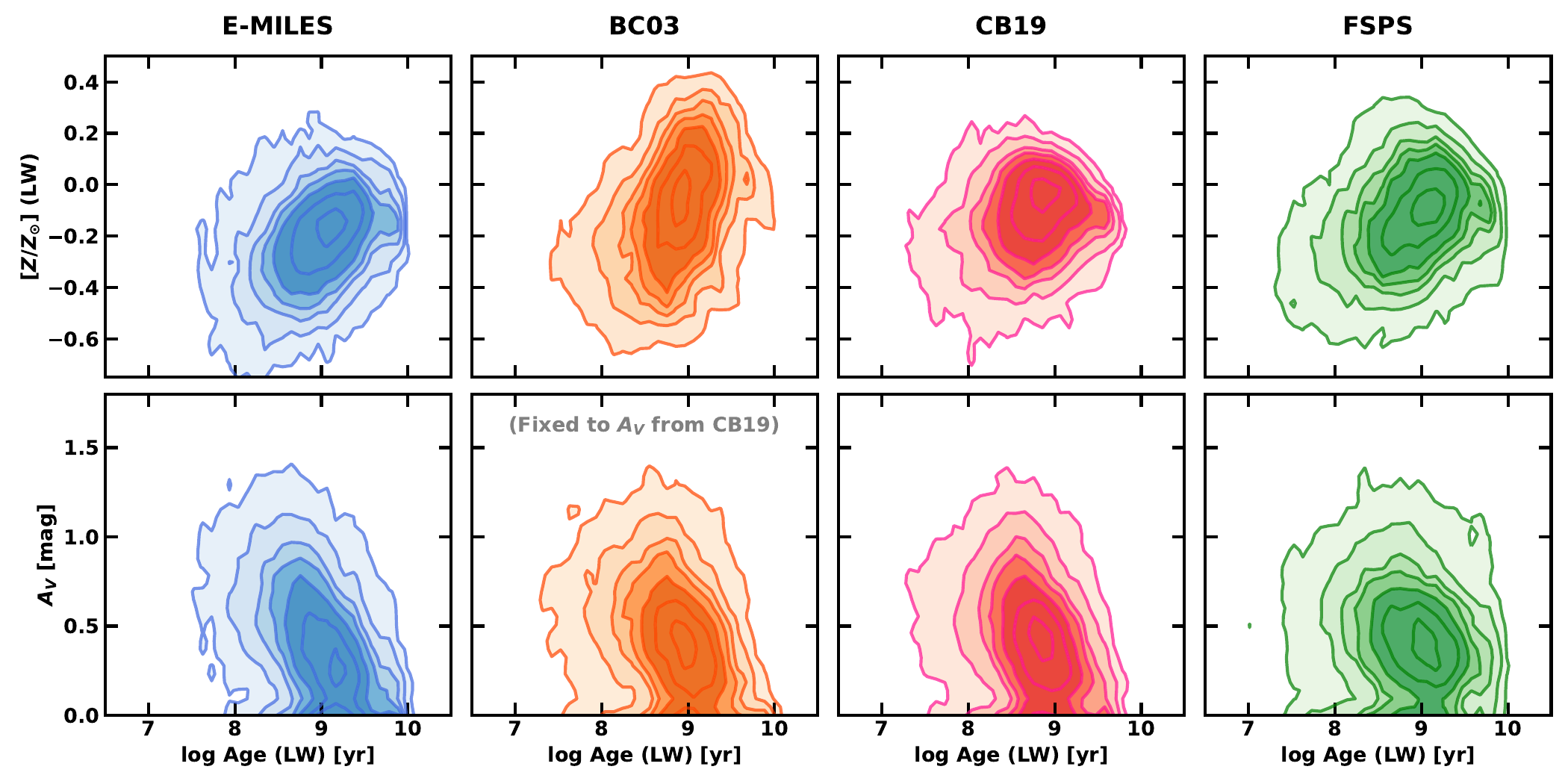}
\caption{
Contour plots showing the distribution of output parameters, including age, metallicity, and $V$-band extinction magnitude ($A_{V}$), derived from the best-fit pPXF results for all spaxels in 19 PHANGS-MUSE galaxies.
In each panel, the contour levels represent the 50th, 70th, 80th, 85th, 90th, 95th, and 99th percentiles of the 2D distributions of spaxel counts.
The top four panels show the distributions of age and metallicity, and the bottom panels show the distributions of age and $A_{V}$.
Results from the four SPS models are displayed in separate columns: E-MILES (first), BC03 (second), CB19 (third), and FSPS (fourth).
Note that for BC03, the $A_{V}$ values are fixed to those derived from the CB19 model in this study.
\label{fig:param}}
\end{figure*}

{\color{blue} \textbf{Figure \ref{fig:comp_stp}}} presents comparisons of luminosity-weighted mean stellar ages (left) and metallicities (right) for all galaxies.
For this figure, digitized data from \citet{pes23} were interpolated to match our radial bins.
The stellar age estimates derived in this work roughly agree with those presented in \citet{pes23} within a scatter of $\sim0.2~{\rm dex}$.
This large scatter is expected because the data from \citet{pes23} represent luminosity-weighted mean values in their radial bins, which are not ideal for pixel-to-pixel comparisons.
Notably, in the regions within $r/R_{\rm e}<2$, the stellar ages from this study tend to be systematically overestimated by $0.1-0.2~{\rm dex}$ compared with those from \citet{pes23}, while this systematic offset appears to diminish in the outer regions ($r/R_{\rm e}>2$) despite a larger scatter.
These findings suggest that the binning process marginally affects the accuracy of stellar ages in the inner part of galaxies, likely due to the mixing of flux from older spaxels with higher surface brightness and younger spaxels with relatively lower surface brightness, which can bias the age estimates toward older values.
In contrast, the radial distribution of metallicity exhibits a good match with that from \citet{pes23} across all radial bins, with a minor scatter of $\sim0.1~{\rm dex}$, revealing that binning has little impact on metallicity measurements.
This may be associated with the flatter gradient of stellar metallicity compared to that of stellar ages in most PHANGS-MUSE galaxies, as also noted in \citet{pes23}.

In conclusion, the SPHEREx binning procedure in this study has negligible or minimal effects on stellar kinematics and metallicities.
This implies that binning does not significantly affect the overall results from full-spectrum fitting, in particular for the resolved stellar mass estimation.
However, in the case of stellar ages, we find a tendency for overestimation 
by $\sim0.1-0.2~{\rm dex}$ with a large scatter in the regions within $r/R_{\rm e}<2$, suggesting a marginal effect of binning on the age distribution.

\section{The Model Dependence of Stellar Population Properties}
\label{sec:model}

\begin{deluxetable*}{ccccccc}
	\tabletypesize{\footnotesize}
	\setlength{\tabcolsep}{0.15in}
	\tablecaption{Percentiles of Stellar Ages, Metallicities, and Internal Extinctions for Each SSP Model}
	\tablehead{\colhead{Parameter} & \colhead{Model} & \colhead{$Q_{2.5\%}$} & \colhead{$Q_{16\%}$} & \colhead{$Q_{50\%}$} & \colhead{$Q_{84\%}$} & \colhead{$Q_{97.5\%}$}}
	\startdata
    ${\rm log~Age}$ (yr)
    & E-MILES & 8.11 & 8.60 & 9.00 & 9.35 & 9.74 \\
    & BC03    & 7.93 & 8.49 & 8.88 & 9.22 & 9.54 \\
    & CB19    & 7.93 & 8.43 & 8.79 & 9.11 & 9.48 \\
    & FSPS    & 7.85 & 8.42 & 8.90 & 9.32 & 9.71 \\ \hline
    ${\rm [Z/Z_{\odot}]}$
    & E-MILES & $-0.52$ & $-0.35$ & $-0.20$ & $-0.08$ & $0.05$ \\
    & BC03    & $-0.49$ & $-0.32$ & $-0.10$ & 0.10 & 0.28 \\
    & CB19    & $-0.39$ & $-0.24$ & $-0.10$ & 0.01 & 0.09 \\
    & FSPS    & $-0.44$ & $-0.29$ & $-0.13$ & 0.02 & 0.16 \\ \hline
    $A_{V}$ (mag)
    & E-MILES & 0.00 & 0.11 & 0.38 & 0.69 & 1.01 \\
    & BC03    & 0.00 & 0.17 & 0.42 & 0.69 & 1.00 \\
    & CB19    & 0.00 & 0.17 & 0.42 & 0.69 & 1.00 \\
    & FSPS    & 0.00 & 0.19 & 0.42 & 0.68 & 1.01 \\
	\enddata
	\label{tab:param}
\end{deluxetable*}

\subsection{Distributions of Ages, Metallicities, and Internal Extinction}
\label{sec:output}

The choice of SPS models can influence the derived stellar populations obtained through full-spectrum fitting \citep{ge19}.
In this section, we compare the stellar properties derived from four SPS models to clarify the systematic effects of model selection.
{\color{blue} \textbf{Figure \ref{fig:param}}} displays the distributions of (luminosity-weighted) stellar ages, metallicities (${\rm [Z/Z_{\odot}]}$), and internal extinction magnitudes ($A_{V}$) for all SPHEREx-binned pixels from 19 PHANGS-MUSE galaxies.
In this figure, different horizontal panels correspond to the four SPS models.
The associated statistics are listed in {\color{blue} \textbf{Table 1}}.

Across all SPS models, the mean stellar ages exhibit broad distributions, ranging from ${\rm log~Age~(yr)}\sim7.9-9.7$, indicating the presence of a considerable amount of young stellar populations in the late-type galaxy sample of PHANGS-MUSE.
Stellar metallicity and internal extinction are correlated with stellar age: metallicity increases while dust extinction decreases as stellar populations become old.
These trends align with conventional concepts of the relationships between stellar populations and galaxy evolution.
Stellar ages and metallicities are known to positively correlate with stellar masses, as the SFH and chemical enrichment are closely related to the stellar mass growth, resulting in mass--age and mass--metallicity relations \citep{tre04, gal05, gon15}.
The negative correlation between internal dust extinction and age is also expected, as dust is more abundant in younger star-forming regions than in older stellar regions.

Remarkable discrepancies in stellar population parameters appear depending on the adopted SPS models.
First, the E-MILES model with the superyoung templates tend to predict older stellar ages and lower metallicities than the other SPS models by $0.1-0.2~{\rm dex}$.
This indicates that results from previous PHANGS-MUSE studies \citep{ems22, pes23} might vary slightly if SPS models other than the E-MILES templates are used.
Second, the metallicities derived using the BC03 model exhibit a much wider distribution than those from the other SPS models, with the 16th and 84th percentile interval being approximately $0.42~{\rm dex}$ for BC03, which is $0.1-0.2~{\rm dex}$ broader than that for the other models.
This wide distribution could be attributed to the tendency of the BC03 model to predict a higher fraction of pixels with super-solar metallicity.
The remaining two SPS models, CB19 and FSPS, yield similar results for age and metallicity.

The $V$-band extinction magnitudes exhibit consistent distributions across the SPS models.
For the BC03 model, however, the extinction magnitudes were fixed to match those from the CB19 model, as represented in Figure \ref{fig:param} and Table 1.
Before this adjustment, we observed that BC03 significantly underestimated the $A_{V}$ values by $\sim0.3~{\rm mag}$ compared with other SPS models, with around $30\%$ of all spaxels showing $A_{V}=0.0$, which appeared unphysical.
As this discrepancy of $A_{V}$ can introduce a potential systematic bias in the other stellar population parameters, we manually adjusted the $A_{V}$ for BC03 throughout this study.

\begin{figure*}
\centering
\includegraphics[width=1.0\textwidth]{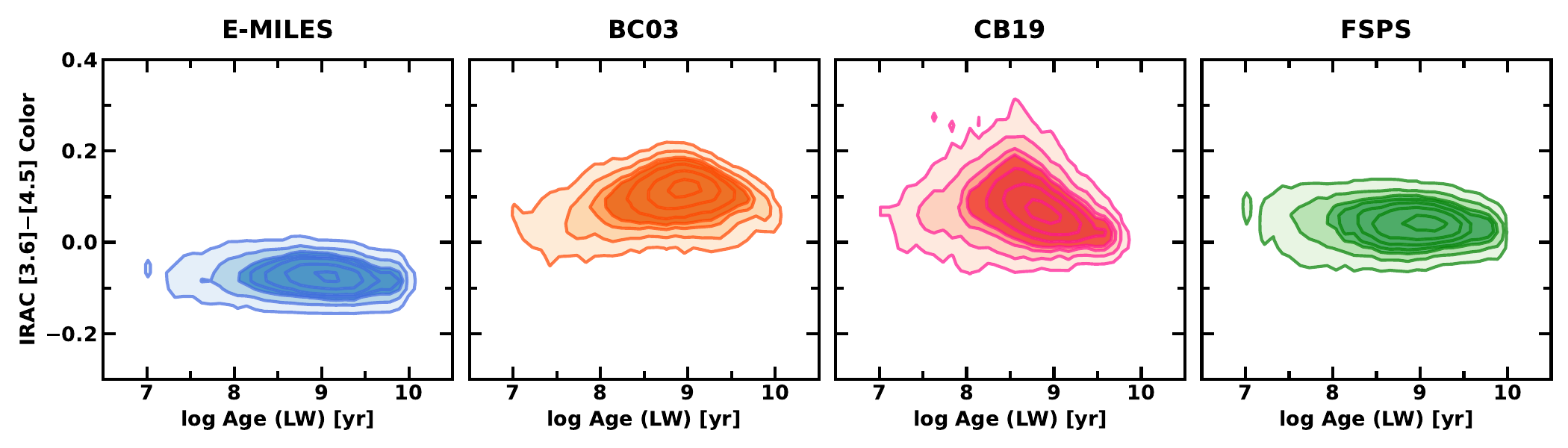}
\caption{
Contour plots showing the distribution of Spitzer/IRAC colors at $3.6~{\rm \mu m}$ and $4.5~{\rm \mu m}$ ([3.6]$-$[4.5]) as a function of stellar ages.
In each panel, colored contours show the stellar-only IRAC color distributions derived from the best-fit spectra of the pPXF results, based on different SPS models: E-MILES (blue), BC03 (orange), CB19 (red), and FSPS (green).
The contour levels represent the 50th, 70th, 80th, 85th, 90th, 95th, and 99th percentiles of the 2D distributions of spaxel counts.
\label{fig:IRcolor}}
\end{figure*}

\subsection{Predictions of IRAC Colors}
\label{sec:IRcolor}

In this section, we explore the distributions of the NIR color index, specifically Spitzer/IRAC $[3.6]-[4.5]$, predicted by four different SPS models.
This analysis is helpful in evaluating the effect of the choice of SPS models on the prediction of stellar properties in the NIR.
Since the $[3.6]-[4.5]$ color index is sensitive to the treatment of TP-AGB stars and molecular absorption features in SPS models, this color serves as a useful diagnostic tool for assessing the NIR SEDs of stellar populations.
Moreover, the $[3.6]-[4.5]$ color has been widely used to trace old stellar populations and disentangle stellar light from non-stellar contributions, such as dust and polycyclic aromatic hydrocarbon (PAH) emissions, in nearby galaxies.
For instance, studies from the ${\rm S^{4}G}$ survey demonstrated that the $[3.6]-[4.5]$ color is effective to derive ``uncontaminated'' stellar masses by distinguishing stellar light from total IRAC $3.6~{\rm \mu m}$ fluxes, with a typical range of $[3.6]-[4.5]$ for old stars between $-0.2$ and $0.0$ \citep{mei14, que15}.
In this context, we examine the systematic differences in $[3.6]-[4.5]$ color distributions across SPS models to evaluate their NIR SED predictions.

\begin{figure*}
\centering
\includegraphics[width=1.0\textwidth]{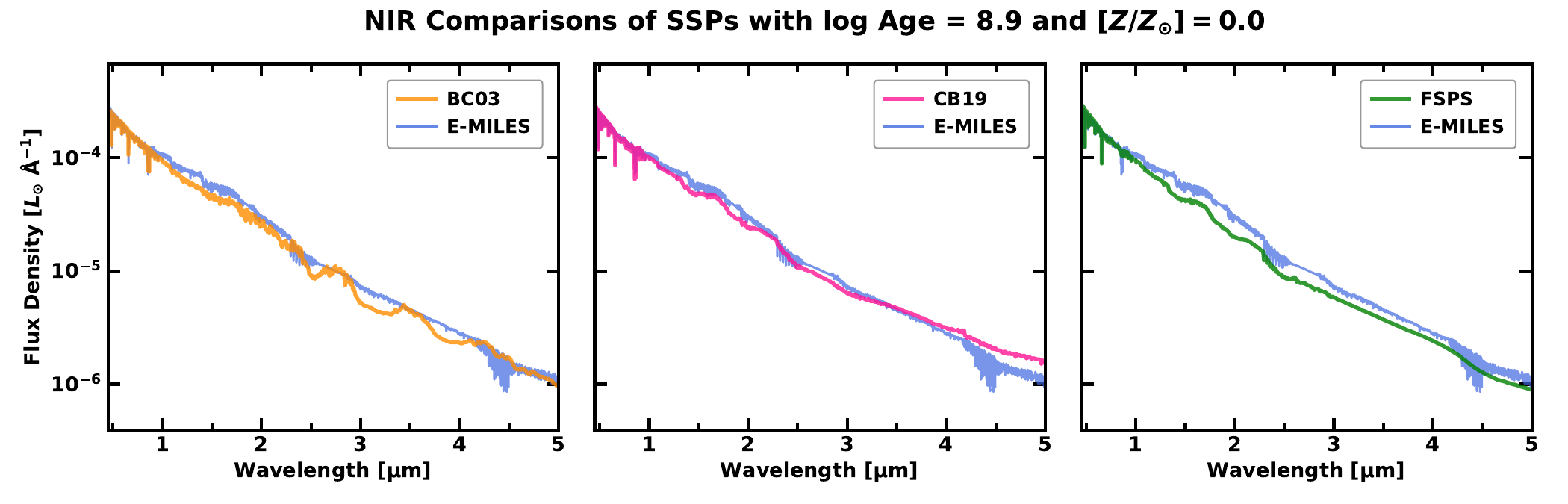}
\caption{
SEDs in the wavelengths of $0.45-5.0~{\rm \mu m}$ from SSP templates, assuming a stellar age of ${\rm log~Age~(yr)}=8.9$ and solar metallicity ($[Z/Z_{\odot}]=0.0$), approximately corresponding to the median values of stellar parameters in {\color{blue} \textbf{Table 1}}.
Each panel compared the SEDs of BC03 (left), CB19 (middle), and FSPS (right) with those from E-MILES, highlighting the differences among the models in the NIR regime.
\label{fig:IRsed}}
\end{figure*}

To highlight these variations, {\color{blue} \textbf{Figure \ref{fig:IRcolor}}} shows the distribution of $[3.6]-[4.5]$ colors as a function of stellar age for each SPS model.
These colors were derived from pPXF fitting by applying IRAC bandpasses to the best-fit stellar templates, considering only stellar emissions.
Among the SPS models, E-MILES exhibits a particularly narrow color distribution between $-0.15$ and $0.0$.
This color range is consistent with the stellar IRAC colors provided in the ${\rm S^{4}G}$ survey results (see Table 1 in \cite{que15}).
In contrast, the other SPS models show substantially redder and broader color distributions, ranging from $-0.05$ to $\sim0.2$.
The age dependence of the $[3.6]-[4.5]$ colors also varies with the SPS models.
While E-MILES and FSPS show nearly no correlation between color and stellar age, BC03 and CB19 exhibit slightly age-dependent trends, with colors becoming the reddest around ${\rm log~Age}=8.5-9.0$.
These discrepancies may be attributed to the variations in the treatment of intermediate-age stellar emission at NIR wavelengths.
Although E-MILES systematically overestimates stellar ages by $0.1-0.2~{\rm dex}$ as presented in Section \ref{sec:output}, this age difference cannot fully explain the substantial color discrepancies between E-MILES and the other SPS models, as the $[3.6]-[4.5]$ color in E-MILES consistently falls between $-0.15$ and $0.0$ across all age ranges.
Thus, these NIR color distributions indicate intrinsic differences in the NIR SEDs among the SPS models.

\begin{figure*}
\centering
\includegraphics[width=1.0\textwidth]{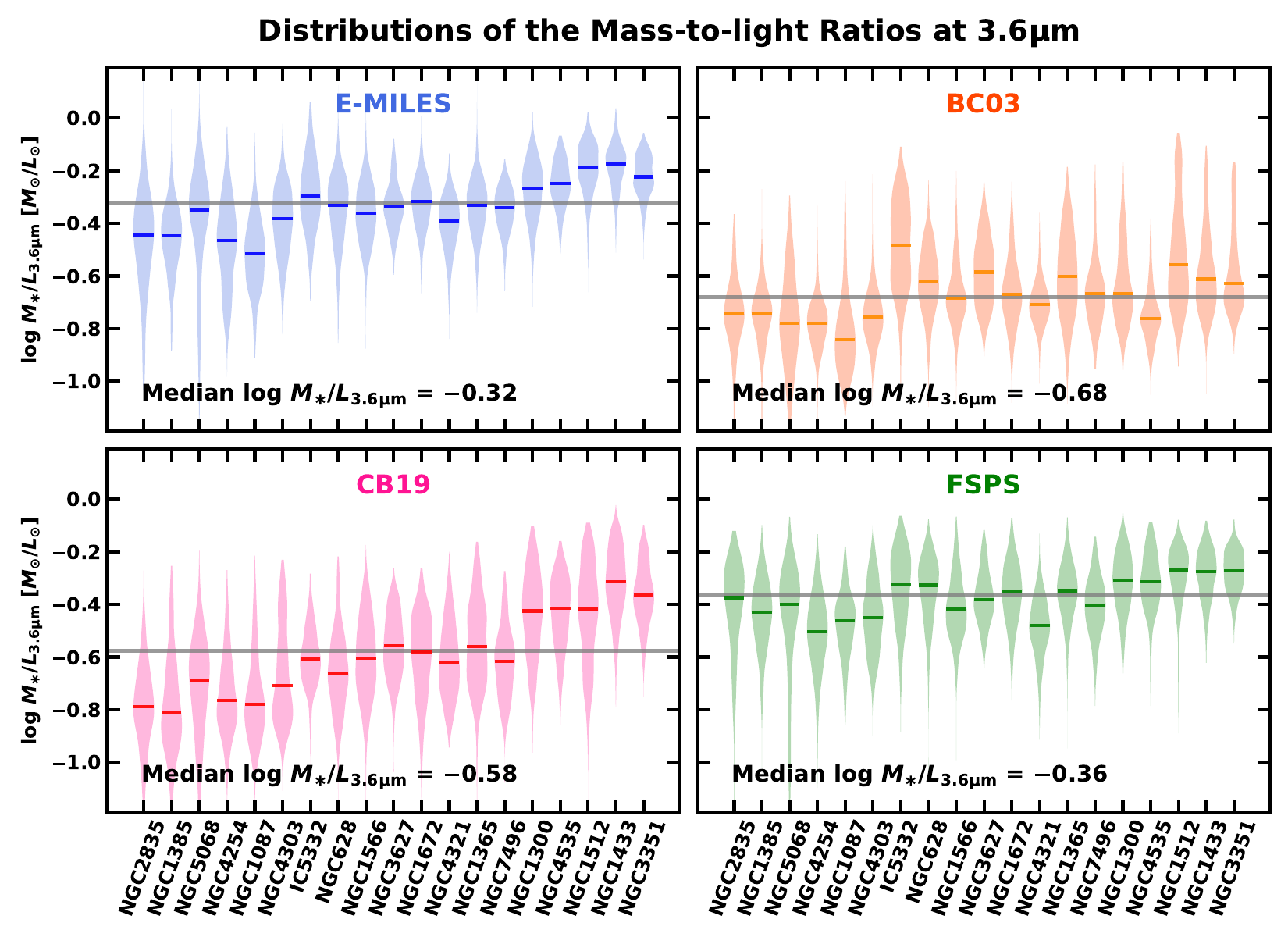}
\caption{
Violin diagrams illustrating the distributions of mass-to-light ratios at IRAC $3.6~{\rm \mu m}$ for each SPS model.
The results are displayed in separate panels: E-MILES (top left), BC03 (top right), CB19 (bottom left), and FSPS (bottom right).
Within each panel, the distributions are presented for all individual galaxies included in this study.
These galaxies in the x-axis are ordered by their median stellar ages derived using E-MILES model.
The median mass-to-light ratios for all galaxies are represented by gray horizontal lines, with the specific values indicated at the bottom of each panel.
\label{fig:violin}}
\end{figure*}

To further examine these differences, we illustrate example SEDs predicted by the four different SPS models with a specific age and metallicity in {\color{blue} \textbf{Figure \ref{fig:IRsed}}}.
The SEDs are modeled with ${\rm log~Age~(yr)}=8.9$ and solar metallicity using each SPS model, which approximately corresponds to the median values of the derived stellar population parameters (see {\color{blue} \textbf{Table 1}}).
This figure clearly reveals that the SED deviations between SPS models are larger in the NIR compared to optical wavelengths ($<0.75~{\rm \mu m}$).
In particular, the CO band at $4.2-4.5~{\rm \mu m}$, originating from late-type ($T_{\rm eff}\lesssim6000~{\rm K}$) giant stars \citep[see Figure 7 in][]{roc15}, is more prominent in the E-MILES template compared with the other SPS templates.
This results in a bluer $[3.6]-[4.5]$ color owing to the reduction of the IRAC $4.5~{\rm \mu m}$ fluxes \citep{pel12}.
In contrast, in the other SPS models, these CO absorption lines are either absent (BC03 and FSPS) or significantly weaker (CB19).
The $[3.6]-[4.5]$ colors from E-MILES align best with observed global $[3.6]-[4.5]$ colors of $-0.2\lesssim[3.6]-[4.5]\lesssim0$ \citep{pah04, pel12}, reflecting the detailed treatment of the CO absorption features.

Variations in the treatment of TP-AGB stars, which dominate NIR emission, also contribute to the differences in NIR SEDs among the SPS models.
BC03, in particular, is known to exhibit considerable differences in TP-AGB emissions with other SPS models \citep{mar06, con09, roc16}.
In addition, CB19 shows a more gradual SED slope at wavelengths longer than $4~{\rm \mu m}$, compared to other SPS models.
These differences in the SED features might contribute to the redder $[3.6]-[4.5]$ colors of BC03 and CB19 shown in {\color{blue} \textbf{Figure \ref{fig:IRcolor}}}.
Meanwhile, FSPS shows similar SED slope to E-MILES across the NIR wavelengths, but they still differ in the CO absorption features, leading to systematic color variations.
Overall, these findings suggest that different SPS models have systematic differences in modeling stellar NIR emissions, which may impact the estimation of stellar mass-to-light ratios in the NIR. 
Future spectrophotometric data from SPHEREx, covering a spectral range of $0.75-5~{\rm \mu m}$ will enable the assessment of the reliability of SPS models and allow for updates to their prescriptions in the NIR range.

\subsection{Dependence of Stellar Mass-to-light Ratios on Ages, Metallicities, and SPS Models}
\label{sec:age}

In this section, we investigate the dependence of mass-to-light ratios in the NIR on stellar population properties and underlying SPS models.
{\color{blue} \textbf{Figure \ref{fig:violin}}} displays the distributions of mass-to-light ratios at IRAC $3.6~{\rm \mu m}$ for all spaxels in 19 PHANGS-MUSE galaxies, derived using four SPS models.
This figure indicates notable differences in mass-to-light ratios among the SPS models.
The mass-to-light ratios derived from the E-MILES and FSPS models are broadly consistent, with FSPS showing a slight underestimation by a median offset of $\sim0.04~{\rm dex}$ ($\sim9\%$).
In contrast, BC03 and CB19 significantly underestimate mass-to-light ratios, with median offsets of $\sim0.36~{\rm dex}$ ($\sim56\%$) and $\sim0.26~{\rm dex}$ ($\sim45\%$), respectively, compared to E-MILES.
These discrepancies are larger than the typical standard deviations of mass-to-light ratios within individual galaxies (median $\sim 0.15~{\rm dex}$), highlighting the critical impact of the underlying SPS models on the derived mass-to-light ratios.

\begin{figure*}
\centering
\includegraphics[width=1.0\textwidth]{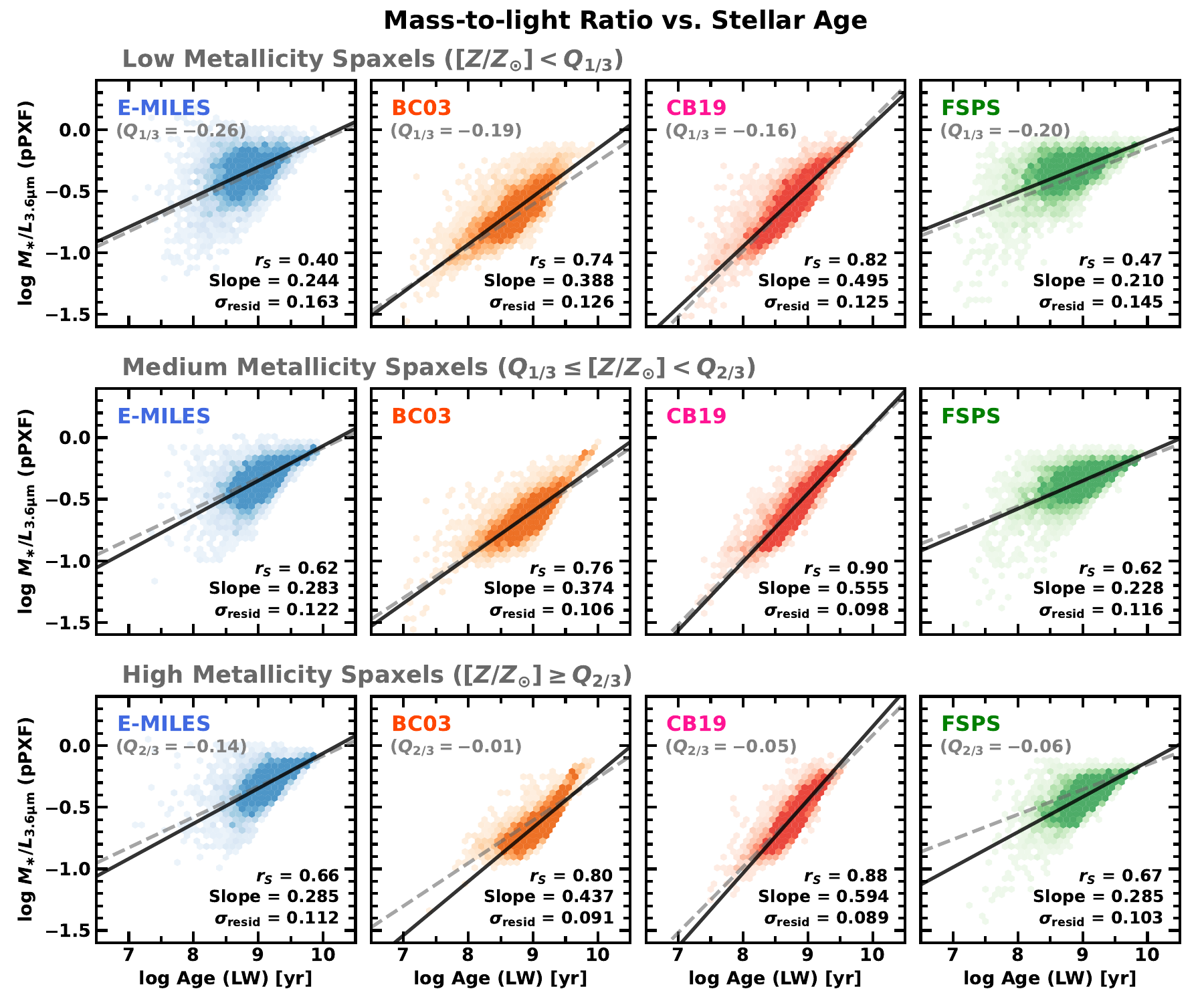}
\caption{
Dependence of the stellar mass-to-light ratio at IRAC $3.6~{\rm \mu m}$ on stellar age, derived using the pPXF method.
Each column displays results from the four SPS models: E-MILES, BC03, CB19, and FSPS (left to right).
Each row represents the age dependence for spaxels grouped by metallicity: low metallicity ($[Z/Z_{\odot}] < Q_{1/3}$), medium metallicity ($Q_{1/3} \leq [Z/Z_{\odot}] < Q_{2/3}$), and high metallicity ($[Z/Z_{\odot}] \geq Q_{2/3}$) from top to bottom.
Here, $Q_{1/3}$ and $Q_{2/3}$ correspond to the 33rd and 67th percentiles of the metallicity distribution from each SPS model.
In each panel, the black solid line indicates the best-fit linear models, with the Spearman's rank correlation coefficients ($r_{S}$), slope, and scatter (standard deviation of the residuals) specified in the bottom-right corner.
The gray dashed line represents the best-fit linear model for all spaxels across all metallicity bins, providing a reference to show deviations caused by metallicity grouping in each panel.
\label{fig:ML_vs_Age}}
\end{figure*}

Systematic variations in mass-to-light ratios are also observed across different galaxies, mainly driven by stellar age, as previously noted by \citet{ems22}.
In {\color{blue} \textbf{Figure \ref{fig:violin}}}, galaxies are organized by their median stellar age, ranging from ${\rm log~Age~(yr) \sim 8.6}$ in NGC 2835 (leftmost) to ${\rm log~Age~(yr) \sim 9.4}$ in NGC 3351 (rightmost).
A clear positive correlation is evident between NIR mass-to-light ratios and stellar ages across all SPS models: galaxies with older stellar populations consistently exhibit higher mass-to-light ratios.
Interestingly, the strength of this correlation varies among the SPS models.
CB19 shows the strongest age dependence, with maximal differences in $M_{\ast}/L_{\rm 3.6\mu m}$ reaching $\sim0.4~{\rm dex}$ between the oldest and youngest galaxies, while E-MILES and FSPS show milder correlations with maximal differences of $\sim0.2~{\rm dex}$.
These findings imply that mass-to-light ratios in the NIR significantly depend on the choice of SPS model and the stellar ages derived from the model.

{\color{blue} \textbf{Figure \ref{fig:ML_vs_Age}}} further examines the relationship between NIR mass-to-light ratios and stellar ages by plotting all spaxel data derived using different SPS models.
This figure is divided vertically into panels based on metallicity bins, defined using the 33rd and 67th percentiles of the metallicity derived from each SPS model.
As expected, a strong positive correlation between stellar ages and mass-to-light ratios is evident across all panels, with Spearman's rank correlation coefficients ($r_{S}$) ranging from $0.4-0.7$ for E-MILES and FSPS, and $0.7-0.9$ for BC03 and CB19.
For old stellar populations (${\rm log~Age~(yr)>9.5}$), the pPXF-derived mass-to-light ratios from each SPS models seem to align well with each other.
However, in regions with young stellar populations, the NIR mass-to-light ratios rapidly decrease to ${\rm log}~M_{\ast}/L_{3.6~{\rm \mu m}}\sim-0.5$ to $-1.0$, with different slopes for each SPS model.
This variation in the age--$M_{\ast}/L_{\rm 3.6 \mu m}$ relation highlights the importance of considering stellar ages and SPS models for stellar mass estimation.
E-MILES and FSPS tend to yield higher stellar masses with weaker correlations with stellar ages compared with other models.
Conversely, BC03 and CB19 tend to predict substantially lower stellar masses in regions with young stellar populations (${\rm log~Age~(yr)}<9$).
The relations derived from BC03 and CB19 exhibit steeper slopes ($\sim0.4-0.6$) with slightly smaller scatters compared to those from E-MILES and FSPS ($\sim0.2-0.3$).
This trend naturally leads to the underestimation of stellar mass-to-light ratio for BC03 and CB19 in younger stellar regions relative to E-MIELS and FSPS.

Stellar metallicity seems to have a minor impact on the age--$M_{\ast}/L_{\rm 3.6 \mu m}$ relation.
Although the slopes may slightly increase in high-metallicity bins across all SPS models, the effect is subtle and less pronounced than the overall scatter of the relation.
In low-metallicity bins, the scatters of the relation increase and the Spearman's correlation coefficients decrease, suggesting that NIR mass-to-light ratios in metal-poor regions are prone to deviate from a simple linear relation with stellar age.
Nonetheless, this metallicity dependence is far less significant than the age dependence in determining stellar mass-to-light ratios, as indicated by much lower correlation coefficients ($r_{S}\lesssim0.1$) for the metallicity--$M_{\ast}/L_{\rm 3.6 \mu m}$ relation compared to the age--$M_{\ast}/L_{\rm 3.6 \mu m}$ relation.
In conclusion, these findings highlight the importance of carefully considering stellar ages to achieve accurate stellar mass estimation using any SPS models, especially in late-type galaxies.

\begin{figure*}
\centering
\includegraphics[width=1.0\textwidth]{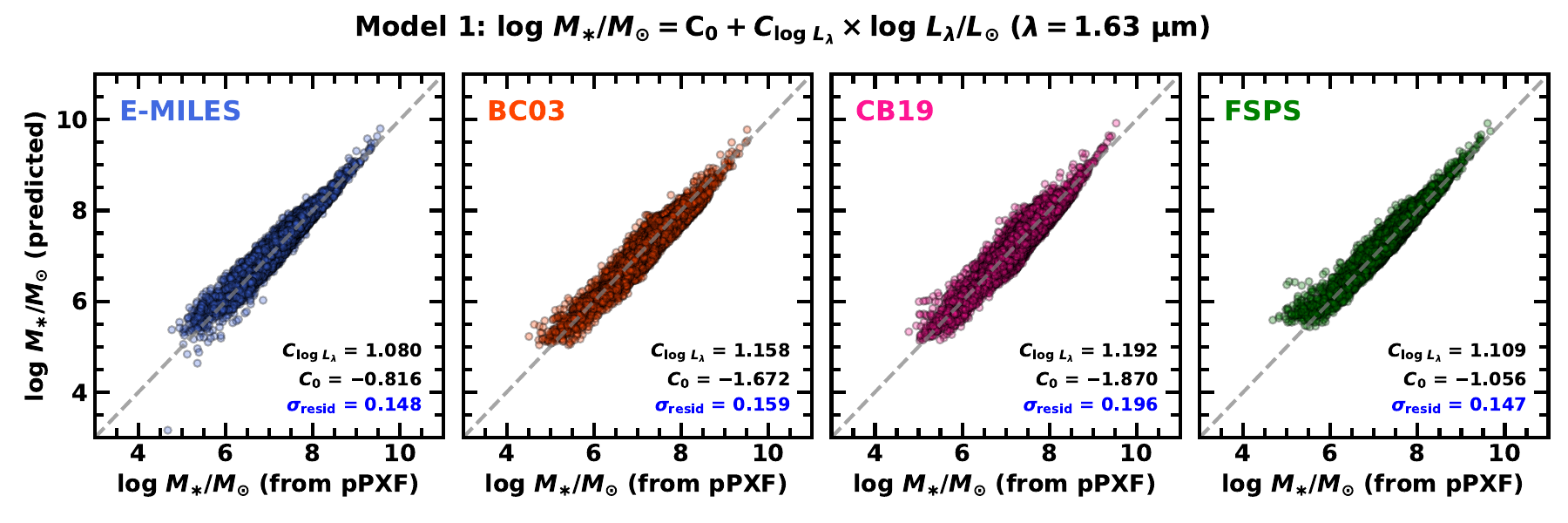}
\caption{
Comparisons of stellar masses obtained from pPXF (x-axis) and values predicted by the best-fit models of a linear relation with specific luminosity (y-axis).
The gray dashed lines represent the one-to-one lines. 
In each panel, the slopes, intercepts, and their respective scatters (standard deviation of the residuals in dex) are noted in the lower right corner.
\label{fig:sigma}}
\end{figure*}

\section{Stellar Mass Estimation with SPHEREx Based on Different SPS Models}
\label{sec:spherex}

One of the primary objectives of this study is to identify the optimal methods for estimating the stellar masses of nearby galaxies in preparation for the all-sky NIR spectrophotometric data that will be provided by the SPHEREx mission.
In this section, we explore strategies for predicting stellar masses using mock SPHEREx data, treating the resolved stellar masses derived from the SPHEREx-binned PHANGS-MUSE data as fiducial values.
For this purpose, we used the monochromatic luminosity and/or colors extracted from SPHEREx wavelength channels.

\begin{figure}
\centering
\includegraphics[width=0.5\textwidth]{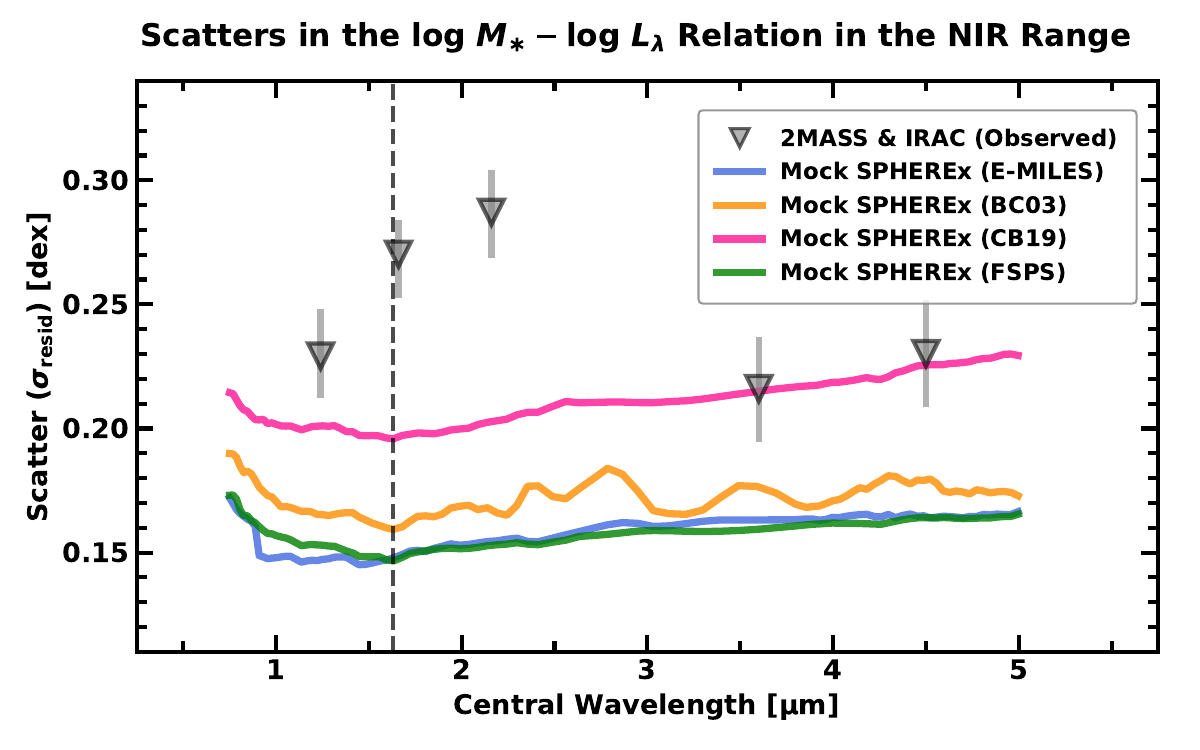}
\caption{
Scatter (standard deviation of the residuals in dex) of the best-fit linear model for the stellar mass (${\rm log}~M_{\ast}$) vs. luminosity relationship at specific wavelengths (${\rm log}~L_{\lambda}$), depending on the central wavelength of photometric bands.
Gray triangles represent the scatters from the relation using observed luminosities from 2MASS $JHK_{s}$ and IRAC $3.6$, $4.5~{\rm \mu m}$.
Error bars indicate the full range of scatters for stellar masses derived from the four SPS models.
The results using luminosities of 96 SPHEREx channels computed by the four SPS models are presented as blue (E-MILES), orange (BC03), brown (CB19), and red (FSPS) curves.
The vertical dashed line indicates the central wavelength of the SPHEREx channel with the minimum mean scatter for the SPS models, which is $1.63~{\rm \mu m}$.
\label{fig:SPxBands}}
\end{figure}

To simulate SPHEREx photometric data for PHANGS-MUSE galaxies, we utilized SEDs at NIR wavelengths obtained from the best-fit results of SPHEREx-binned pixels through full-spectrum fitting.
From these NIR SEDs, we extracted fluxes at all SPHEREx wavelength channels.
As the filter transmission curves for SPHEREx channels have not been publicly available yet, we created mock SPHEREx channels with Gaussian-shaped filters, following a similar approach to \citet{kim24}.
These mock channels were centered on 96 wavelengths listed in the SPHEREx Public Github repository\footnote[9]{Available in \url{https://github.com/SPHEREx/Public-products}}. 
The effective widths of the SPHEREx bands were determined based on the spectral resolutions provided by \citet{dor18}: $\lambda/\Delta\lambda_{\rm FWHM}=41$ for $0.75-2.42~{\rm \mu m}$, $\lambda/\Delta\lambda_{\rm FWHM}=35$ for $2.42-3.82~{\rm \mu m}$, $\lambda/\Delta\lambda_{\rm FWHM}=110$ for $3.82-4.42~{\rm \mu m}$, and $\lambda/\Delta\lambda_{\rm FWHM}=130$ for $4.42-5.00~{\rm \mu m}$.
Using these mock SPHEREx filter responses, we calculated the corresponding fluxes within the NIR SEDs for the four NIR-covering SPS models, using the \texttt{PyPhot} package.

To evaluate the accuracy of stellar mass estimation based solely on monochromatic luminosity at a SPHEREx spectral channel, we developed a simple model to predict fiducial stellar masses from specific luminosities at each SPHEREx channel: ${\rm log~}M_{\ast}/M_{\odot}=C_{0}+C_{{\rm log}L_{\lambda}}\times{\rm log}~L_{\lambda}/L_{\odot}$ (Model 1).
{\color{blue} \textbf{Figure \ref{fig:sigma}}} compares the stellar masses predicted using pPXF and Model 1 at $1.63~{\rm \mu m}$ for the four SPS models, covering a mass range from $10^{4}~M_{\odot}$ to $10^{10}~M_{\odot}$.
We applied this model to all SPHEREx channels and determined the best-fit parameters of $C_{0}$ and $C_{{\rm log}L_{\lambda}}$.
Subsequently, we examined the scatter (standard deviation of the residuals) in the model across all SPHEREx channels, as shown in {\color{blue} \textbf{Figure \ref{fig:SPxBands}}}.
This figure displays the scatter trends over the SPHEREx wavelength range for the four SPS models.
The gray triangles represent the scatter when using the observed broad-band luminosities for Model 1, which were measured from the reprojected 2MASS $JHK_{s}$ and Spitzer/IRAC $3.6~{\rm \mu m}$ and $4.5~{\rm \mu m}$ images.
Notably, the scatter in the ${\rm log}~M_{\ast}$ vs. ${\rm log}~L_{\lambda}$ relations is lower for mock SPHEREx luminosities across all SPS models compared with the observed broad-band data, indicating that SPHEREx luminosity is more suitable for robust stellar mass measurements than conventional broad-band photometry.

\begin{figure*}
\centering
\includegraphics[width=1.0\textwidth]{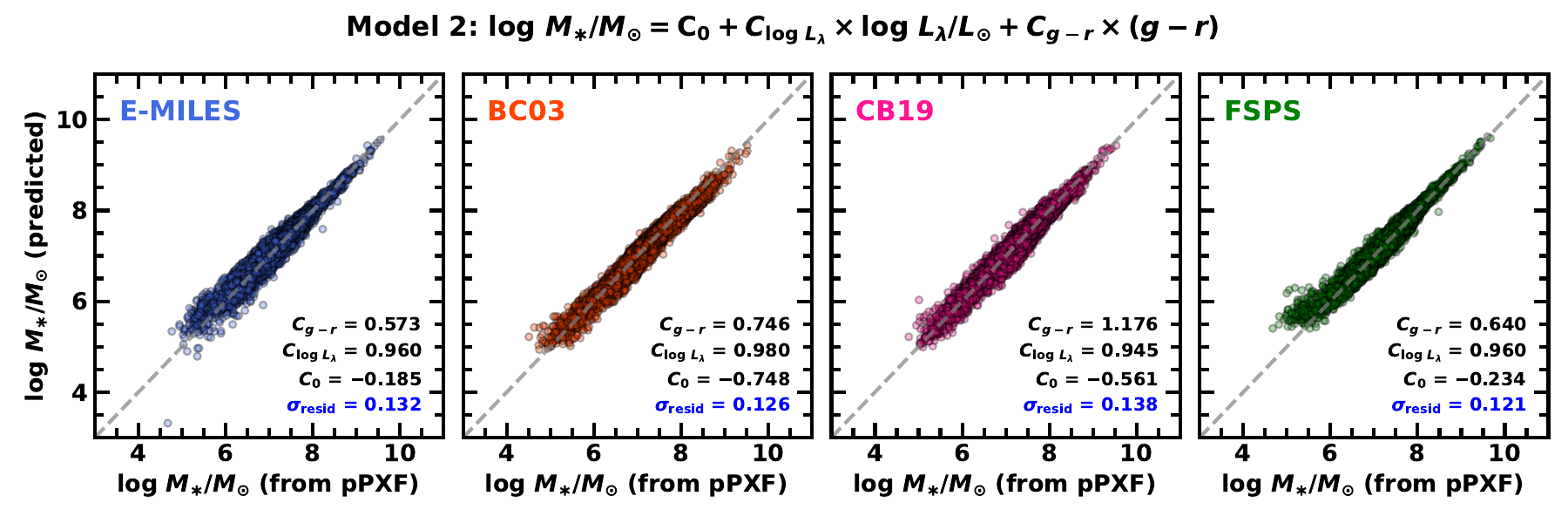}
\caption{
Comparisons of stellar masses obtained from the pPXF (x-axis) with the values predicted by the best-fit models with linear combinations of specific luminosity and SDSS $g-r$ color (y-axis).
The panels are organized similarly to those in {\color{blue} \textbf{Figure \ref{fig:sigma}}}.
The texts at the lower right corner of each panel indicate the scatters of models and best-fit coefficients, including the slopes of specific luminosity ($C_{{\rm log}L_{\lambda}}$) and SDSS $g-r$ color ($C_{g-r}$).
\label{fig:sigma_MLR}}
\end{figure*}

For Model 1, the minimum scatter consistently occurs at approximately $1.63~{\rm \mu m}$, almost independent of the adopted SPS models, suggesting that this wavelength is optimal for robust stellar mass estimation.
This finding is consistent with the conclusion of \citet{kim24}, which explored the distributions of stellar mass-to-light ratios in NIR wavelengths of nearby galaxies using broad-band SED fitting.
The preference for $1.63~{\rm \mu m}$ may be attributed to the spectral bump caused by the ${\rm H^{-}}$ ion from cool stars \citep{saw02, sor10}, which appears universally in SPS model templates, making it suitable for stellar mass estimations.
The scatter for SPS models ranges from $0.15~{\rm dex}$ to $0.20~{\rm dex}$, with the smallest and largest values corresponding to E-MILES and CB19, respectively, as shown in {\color{blue} \textbf{Figure \ref{fig:sigma}}}.

\begin{figure*}
\centering
\includegraphics[width=1.0\textwidth]{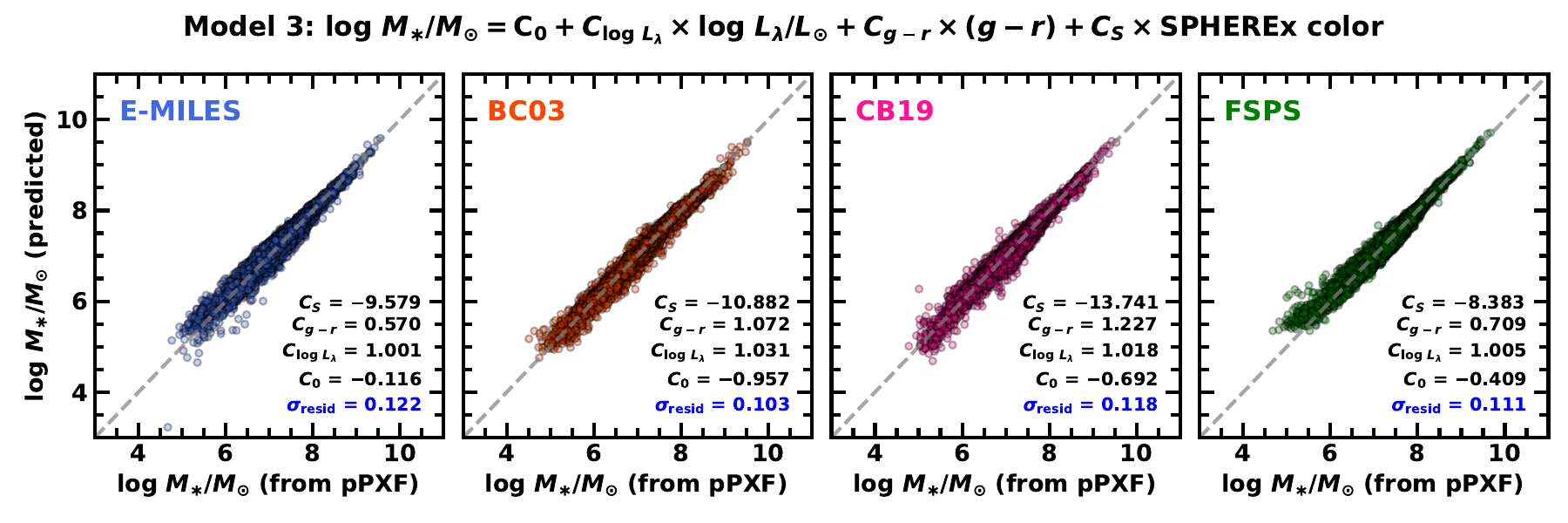}
\caption{
Comparisons of stellar masses obtained from pPXF (y-axis) with the values predicted by the best-fit models with linear combinations of specific luminosity, SDSS $g-r$ color, and SPHEREx color between the $1.14~{\rm \mu m}$ and $1.17~{\rm \mu m}$ bands (x-axis).
The panels are organized similarly to those in {\color{blue} \textbf{Figure \ref{fig:sigma}}}.
The scatters and best-fit coefficients of models, including the slopes of specific luminosity ($C_{{\rm log}L_{\lambda}}$), SDSS $g-r$ color ($C_{g-r}$), and the SPHEREx color ($C_{S}$), are presented on the bottom right of each panel.
\label{fig:sigma_MLR2}}
\end{figure*}

To further refine the mass prediction models, we added color terms.
Conventionally, the color index has been widely used as a secondary parameter to improve the accuracy of stellar mass estimation because it is closely related to the stellar population and mass-to-light ratio.
We examined the influence of including color by adopting a $g-r$ color ({\color{blue} \textbf{Figure \ref{fig:sigma_MLR}}}; Model 2) and an NIR color derived from mock SPHEREx data ({\color{blue} \textbf{Figure \ref{fig:sigma_MLR2}}}; Model 3).
The motivation for using the $g-r$ color is that these complementary data will be publicly available through the Pan-STARRS \citep{cha16}, DESI Legacy Imaging Surveys \citep{dey19}, Dark Energy Survey \citep{abb18, abb21}, and Legacy Survey of Space and Time \citep[LSST;][]{ive19}, covering the entire sky.
We measured $g-r$ colors in all SPHEREx-binned pixels using their best-fit templates derived from pPXF fitting.
{\color{blue} \textbf{Figure \ref{fig:sigma_MLR}}} shows that the SDSS $g-r$ color effectively reduces scatter to $0.12-0.14~{\rm dex}$ for all SPS models, with a dramatic decrease for the CB19 model from $0.20~{\rm dex}$ to $0.14~{\rm dex}$.

\begin{deluxetable*}{ccccc}
	\tabletypesize{\footnotesize}
	\setlength{\tabcolsep}{0.25in}
	\tablecaption{Parameters and Scatters of the Models for Scaling Relations}
	\tablehead{\colhead{Parameter} & \colhead{E-MILES} & \colhead{BC03} & \colhead{CB19} & \colhead{FSPS}}
	\startdata
    \textbf{Model 1} \\ 
    $C_{0}$ & $-0.816$ & $-1.672$ & $-1.870$ & $-1.056$ \\
    $C_{{\rm log}L_{\lambda}}$ & 1.080 & 1.158 & 1.192 & 1.109 \\
    $\sigma_{\rm resid}$ & 0.148 & 0.159 & 0.196 & 0.147 \\ \hline
    \textbf{Model 2} \\ 
    $C_{0}$ & $-0.185$ & $-0.748$ & $-0.561$ & $-0.234$ \\
    $C_{{\rm log}L_{\lambda}}$ & 0.960 & 0.980 & 0.945 & 0.960 \\
    $C_{g-r}$ & 0.573 & 0.746 & 1.176 & 0.640 \\
    $\sigma_{\rm resid}$ & 0.132 & 0.126 & 0.138 & 0.121 \\ \hline  
    \textbf{Model 3} \\ 
    $C_{0}$ & $-0.116$ & $-0.957$ & $-0.692$ & $-0.409$ \\
    $C_{{\rm log}L_{\lambda}}$ & 1.001 & 1.031 & 1.018 & 1.005 \\
    $C_{g-r}$ & 0.570 & 1.072 & 1.227 & 0.709 \\
    $C_{S}$ & $-9.579$ & $-10.882$ & $-13.741$ & $-8.383$ \\
    $\sigma_{\rm resid}$ & 0.122 & 0.103 & 0.118 & 0.111 \\ \hline
	\enddata
	\label{tab:param2}
	\tablenotetext{}{\textbf{Notes.}}
	\tablenotetext{}{\textbf{Model 1:} ${\rm log}~M_{\ast}/M_{\odot}$ $=$ $C_{0}+C_{{\rm log}L_{\lambda}}\times{{\rm log}~L_{\lambda}/L_{\odot}}$ ($\lambda=1.63~{\rm \mu m}$)}
	\tablenotetext{}{\textbf{Model 2:} ${\rm log}~M_{\ast}/M_{\odot}$ $=$ \textbf{Model 1} $+$ $C_{g-r}\times(g-r)$}
	\tablenotetext{}{\textbf{Model 3:} ${\rm log}~M_{\ast}/M_{\odot}$ $=$ \textbf{Model 2} $+$ $C_{S}\times(m_{1.14~{\rm \mu m}}-m_{1.17~{\rm \mu m}})$}
\end{deluxetable*}

Incorporating SPHEREx colors into the mass prediction model further reduces scatter, as shown in {\color{blue} \textbf{Figure \ref{fig:sigma_MLR2}}}.
We tested all combinations of two SPHEREx channels with NIR wavelengths $\lambda<2.4~{\rm \mu m}$ to minimize potential contamination from non-stellar emissions at longer wavelengths.
On average, the optimal SPHEREx color combination that minimizes scatter includes the SPHEREx bands centered at $1.14~{\rm \mu m}$ and $1.17~{\rm \mu m}$, which are adjacent wavelength channels.
Using the optimal SPHEREx color set using the $1.14~{\rm \mu m}$ and $1.17~{\rm \mu m}$ bands, the scatter for all SPS models decreases to $0.10-0.12~{\rm dex}$, significantly improving the results compared with Model 1 shown in {\color{blue} \textbf{Figure \ref{fig:sigma}}}.
These NIR wavelengths do not overlap with any known galaxy emission lines \citep{mart13, izo16}, which indicates that this SPHEREx color set is well-suited for stellar mass estimation.
Although the future SPHEREx data may not provide fluxes exactly at $1.14~{\rm \mu m}$ and $1.17~{\rm \mu m}$, due to uncertainties in the mock SPHEREx filters used in this study and the data processing with the linear variable filters on SPHEREx \citep{cri20}, wavelength channels within the $1.10-1.27~{\rm \mu m}$ range (before the Paschen-${\rm \beta}$ line of $1.28~{\rm \mu m}$) consistently yield lower scatters than other channels.
This suggests that these NIR wavelengths will be practically useful for tracing stellar masses across all SPS models.

The best-fit parameters and scatters of the scaling relations are listed in {\color{blue} \textbf{Table 2}}.
These new scaling relations are based on the fiducial stellar masses derived from spectral fitting, which accounts for variations in stellar populations of late-type galaxies.
Thus, these relations derived from the mock SPHEREx data can be applied to estimate the resolved stellar masses of nearby galaxies using the future SPHEREx dataset.
Furthermore, we accounted for the systematic differences in stellar mass measurement introduced by the choice of the SPS model.
These proactive scaling relations will be valuable for measuring the stellar masses of nearby galaxies in the forthcoming SPHEREx era.

\section{Summary}
\label{sec:summary}

This study aims to develop a preemptive strategy for estimating the stellar masses of nearby galaxies in preparation for the upcoming SPHEREx mission, with two specific objectives:
First, we examine the dependence of SPS models on the derived stellar population properties including resolved stellar masses, using full-spectrum fitting (pPXF) results from 19 PHANGS-MUSE galaxies.
We use the SPHEREx-binned datacube of these galaxies to estimate stellar masses, employing four NIR-covering SPS models: E-MILES, BC03, CB19, and FSPS.
Second, we propose new scaling relations for stellar mass estimation based on mock SPHEREx spectrophotometric data, derived from the best-fit spectral templates from the full-spectrum fitting.
These scaling relations use monochromatic luminosity and/or colors from SDSS optical bands and SPHEREx wavelength channels, offering a tailored approach to measure resolved stellar masses of nearby galaxies with SPHEREx capabilities.
Our main findings are summarized as follows.

\begin{enumerate}
    \item Different SPS models lead to systematic variations in the derived stellar population properties in relation to stellar mass.
    Among the four SPS models, E-MILES predicts higher stellar ages and lower metallicities, while BC03 shows a broader metallicity distribution.
    CB19 and FSPS produce consistent distributions of stellar population parameters.
    \item The predicted SEDs of the four SPS models exhibit notable differences in the NIR, as indicated by the distributions of the $[3.6]-[4.5]$ color.
    E-MILES predicts $[3.6]-[4.5]$ colors between $-0.15$ and $0.0$ across all ages, which are lower than those of other SPS models, due to the prominent CO absorption features in its spectral template.
    In addition, substantial differences exist in spectral features and shapes among the SPS models, which can be effectively resolved by the forthcoming SPHEREx mission.
    \item We find a significant variations in mass-to-light ratios at $3.6~{\rm \mu m}$ (${\rm log}~M_{\ast}/L_{\rm 3.6\mu m}$) among the SPS models.
    E-MILES (median ${\rm log}~M_{\ast}/L_{\rm 3.6\mu m}=-0.32$) and FSPS (median ${\rm log}~M_{\ast}/L_{\rm 3.6\mu m}=-0.36$) yield systematically higher mass-to-light ratios compared to BC03 (median ${\rm log}~M_{\ast}/L_{\rm 3.6\mu m}=-0.68$) and CB19 (median ${\rm log}~M_{\ast}/L_{\rm 3.6\mu m}=-0.58$).
    These discrepancies primarily arise from variations in the age--$M_{\ast}/L_{\rm 3.6\mu m}$ relation across the SPS models.
    BC03 and CB19 exhibit steeper slopes of this relation ($\sim0.4-0.6$) compared to E-MILES and FSPS ($\sim0.2-0.3$), leading to their underestimation of mass-to-light ratio in younger stellar relation relative to E-MILES and FSPS.
    Metallicity has little impact on the age--$M_{\ast}/L_{\rm 3.6\mu m}$ relation for all SPS models. 
    \item We develop new scaling relations for estimating stellar masses using simulated SPHEREx data, by treating our spectrum-derived masses as fiducial values.
    Across all SPS models, the SPHEREx channel at $1.63~{\rm \mu m}$ is identified as the optimal wavelength for accurate mass predictions with minimal scatter ($0.15-0.20~{\rm dex}$).
    Incorporating additional color information, such as the SDSS $g-r$ color and the SPHEREx color between the channels of $1.14~{\rm \mu m}$ and $1.17~{\rm \mu m}$, further reduces scatter to $\sim0.10-0.12~{\rm dex}$.
    These scaling relations effectively account for variations in stellar populations and systematic differences across SPS models, improving mass prediction accuracy.
\end{enumerate}

By combining SPHEREx-released data with archival spectroscopic datasets in the near future, we anticipate being able to effectively validate the scaling relations in the NIR and address the degeneracies between age, metallicity, and extinction that complicate accurate stellar mass estimation.
In addition, our findings suggest the systematic differences in mass-to-light ratios driven by the choice of SPS model, implying the importance of addressing these variations to improve stellar mass estimations in the NIR.

\clearpage

\begin{acknowledgments}
We are grateful to the anonymous referee for constructive comments and suggestions that greatly improved our manuscript.
This work was supported by the National Research Foundation of Korea (NRF) grant funded by the Korean government (MSIT; Nos. 2022R1A4A3031306, 2023R1A2C1006261, and RS-2024-00347548) and Basic Science Research Program through the National Research Foundation of Korea (NRF) funded by the Ministry of Education (No. RS-2024-00452816).
J.H.L. was supported by 
Ascending SNU Future Leader Fellowship through Seoul National University (No. 0409-20240178).
T.K. acknowledges support from the Basic Science Research Program through the National Research Foundation of Korea (NRF) funded by the Ministry of Education (No. RS-2023-00240212).
H.S. acknowledges support from the National Research Foundation of Korea (NRF) grant funded by the Korea government (MSIT; No. RS-2024-00349364). 
H.S.H. acknowledges the support of Samsung Electronic Co., Ltd. (Project Number IO220811-01945-01), the National Research Foundation of Korea (NRF) grant funded by the Korea government (MSIT), NRF-2021R1A2C1094577, and Hyunsong Educational \& Cultural Foundation.
L.C.H. was supported by the National Science Foundation of China (11991052, 12233001), the National Key R\&D Program of China (2022YFF0503401), and the China Manned Space Project (CMS-CSST-2021-A04, CMS-CSST-2021-A06).
D.K. acknowledges the support of the National Research Foundation of Korea (NRF) grant (No. 2021R1C1C1013580).
Based on observations taken as part of the PHANGS-MUSE large program (Emsellem et al. 2022).
Based on data products created from observations collected at the European Organisation for
Astronomical Research in the Southern Hemisphere under ESO programme(s) 1100.B-0651, 095.C-0473,
and 094.C-0623 (PHANGS-MUSE; PI Schinnerer), as well as 094.B-0321 (MAGNUM; PI Marconi),
099.B-0242, 0100.B-0116, 098.B-0551 (MAD; PI Carollo) and 097.B-0640 (TIMER; PI Gadotti). This
research has made use of the services of the ESO Science Archive Facility.
This research has made use of the NASA/IPAC Infrared Science Archive, which is funded by the National Aeronautics and Space Administration and operated by the California Institute of Technology.
\end{acknowledgments}

\software{Numpy \citep{har20}, Matplotlib \citep{hun07}, Scipy \citep{vir20}, Astropy \citep{ast13, ast18, ast22}, pPXF \citep{cap23}, PyPhot \citep{fou22}, extinction \citep{bar16}}

\facility{VLT/MUSE, IRSA, Spitzer, WISE, 2MASS}

\clearpage

\appendix

\section{Comparisons of Stellar Masses with Previous NIR-derived Measurements}
\label{app:A}

\restartappendixnumbering

In this appendix, we compare our fiducial stellar masses, as described in {\color{blue} \textbf{Section \ref{sec:Mstar}}}, with conventional NIR-derived stellar masses to assess the reliability of our estimates.
Stellar masses for large galaxy samples have been determined using NIR luminosities combined with NIR colors \cite[e.g.,][]{mei14, que15, jar23}.
For this comparison, we made use of auxiliary NIR imaging data, including Spitzer/IRAC and WISE, which were previously used to provide the ${\rm S^{4}G}$ stellar mass maps \citep[][hereafter Q15]{que15} and the WISE scaling relations \citep[][hereafter J23]{jar23}, respectively.
It is challenging to directly compare our spectrum-derived stellar masses with those from the Q15 or J23 methods, in the sense that spectroscopy naturally offers a more detailed assessment of stellar population properties in relation to stellar mass (e.g., ages and metallicities), compared to the approach with NIR photometry.
Nonetheless, such comparisons remain valuable for validating spectrum-derived stellar masses and gaining further insights into stellar mass estimation methods.
Acknowledging the inherent limitations of NIR photometric methods, we conducted a comparative analysis to supplement our results in this appendix.

\subsection{The ${\rm S^{4}G}$ Survey}
\label{app:S4G}

\subsubsection{Stellar Light Maps from ${\rm S^{4}G}$}
\label{app:data_S4G}

We obtained the ${\rm S^{4}G}$ data from the NASA/IPAC Infrared Science Archive (IRSA)\footnote[10]{\url{https://irsa.ipac.caltech.edu/data/SPITZER/S4G/galaxies}}.
Specifically, we retrieved the $3.6~{\rm \mu m}$ and $4.5~{\rm \mu m}$ flux maps at the Pipeline 1 (P1) stage, along with the ICA-derived stellar light maps in the $3.6~{\rm \mu m}$ band at the Pipeline 5 (P5) stage.
We created background-subtracted flux maps from the P1 flux maps by determining the background levels using the \textsc{SExtractor} algorithm for background measurement \citep{ber96}.
Subsequently, we generated the ${\rm S^{4}G}$ stellar mass maps following Equation 6 from Q15 and assuming a constant mass-to-light ratio of $M_{\ast}/L_{\rm 3.6\mu m}=0.6$ \citep{mei14}.
Similar to the PHANGS-MUSE datacubes, we resampled the ${\rm S^{4}G}$ maps from their original scale of $0\farcs75~{\rm pixel^{-1}}$ to $6\farcs2~{\rm pixel^{-1}}$ and reprojected them to align with the rebinned PHANGS-MUSE data, using the \texttt{reproject} package\footnote[11]{\url{https://github.com/astropy/reproject}}.
During this process, we applied the P5 ICA-mask maps to remove contamination from foreground stars.
These masks were also applied when comparing the resolved stellar masses with the full-spectrum fitting results from the PHANGS-MUSE data.
Since 5 of the 19 PHANGS-MUSE galaxies (IC 5332, NGC 1433, NGC 1512, NGC 2835, and NGC 7496) lack the P5 stellar light maps from ${\rm S^{4}G}$, we restricted our comparison of spectrum-derived masses with Q15 to the remaining 14 PHANGS-MUSE galaxies.

\subsubsection{Comparison with Q15 Method}
\label{app:comp_S4G}

For the ${\rm S^{4}G}$ galaxies, the Q15 method separated observed IRAC $3.6~{\rm \mu m}$ and $4.5~{\rm \mu m}$ fluxes into stellar and non-stellar components to account for contamination from dust continuum and PAH emissions in the IRAC fluxes.
The ${\rm S^{4}G}$ studies used Independent Component Analysis (ICA) techniques based on the [3.6]$-$[4.5] color information \citep{mei12}, with the assumption of BC03 model and Chabrier IMF \citep{cha03}.
Then, stellar mass-to-light ratios in the $3.6~{\rm \mu m}$ band were assumed to be constant at $M_{\ast}/L_{\rm 3.6\mu m}=0.6$, due to their minimal variations in old stellar populations with ages of $2-12~{\rm Gyr}$ \citep{mei14}.
This approach produced stellar mass maps that were corrected for non-stellar emission contamination for more than a thousand galaxies of the ${\rm S^{4}G}$ sample.
Through this decomposition of stellar and non-stellar light, the ${\rm S^{4}G}$ stellar mass maps can be effectively compared with full-spectrum fitting results, as both methods consider only stellar emissions.

\begin{figure}
\centering
\includegraphics[width=1.0\textwidth]{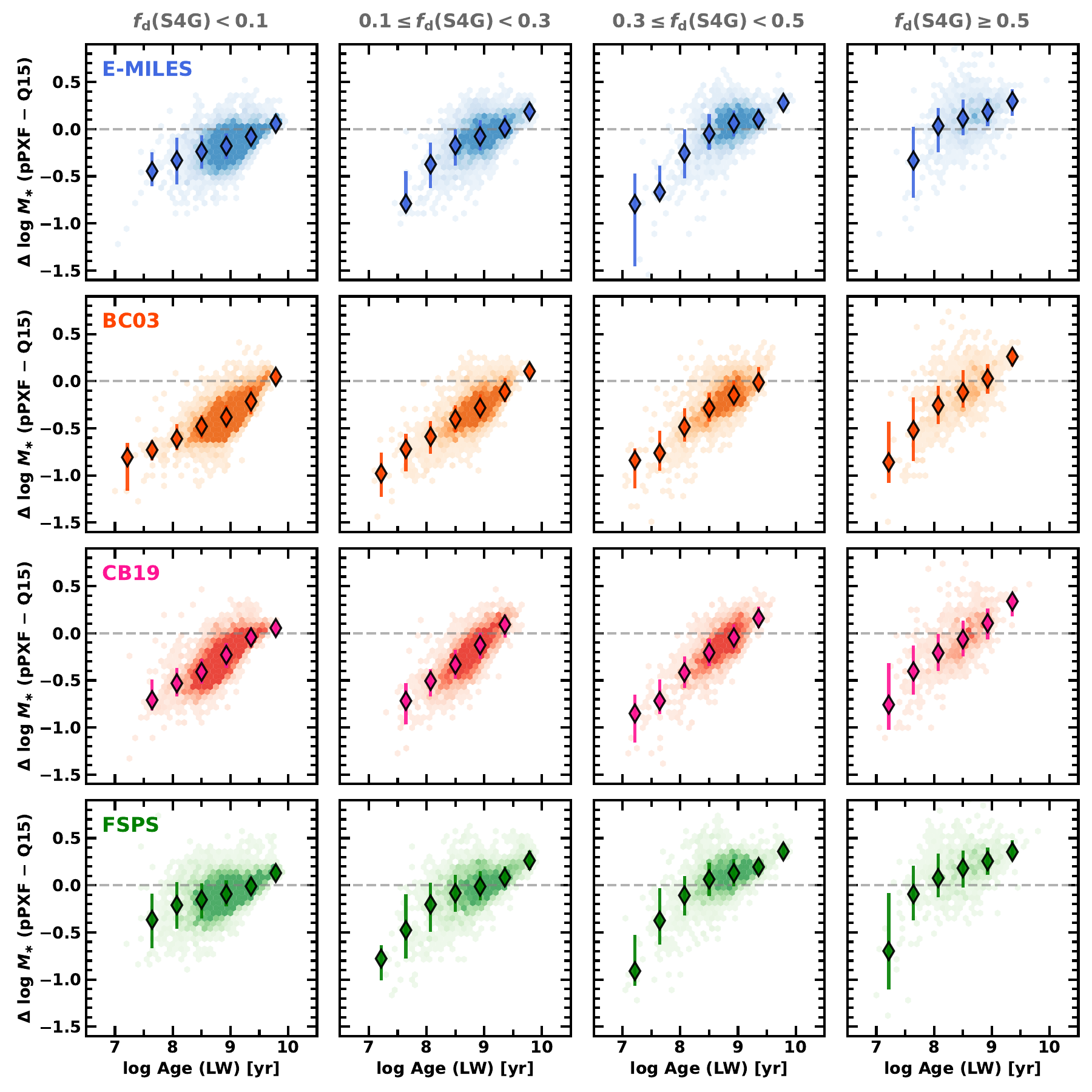}
\caption{
Logarithmic stellar mass differences between our pPXF results and the ${\rm S^{4}G}$ survey, as a function of age.
The panels in this figure are divided by SPS models (vertically) and dust fractions from the ${\rm S^{4}G}$ survey ($f_{\rm d}({\rm S^{4}G})$; horizontally).
The four panels (from top to bottom) show the mass differences for E-MILES, BC03, CB19, and FSPS.
Each vertical column represents a distinct bin of $f_{\rm d}({\rm S^{4}G})$: $f_{\rm d}({\rm S^{4}G})<0.1$, $0.1 \leq f_{\rm d}({\rm S^{4}G})<0.3$, $0.3 \leq f_{\rm d}({\rm S^{4}G})<0.5$, and $f_{\rm d}({\rm S^{4}G})\geq0.5$.
In each panel, median values for age bins are indicated by diamonds.
\label{fig:dMass_vs_Age}}
\end{figure}

\begin{figure}
\centering
\includegraphics[width=1.0\textwidth]{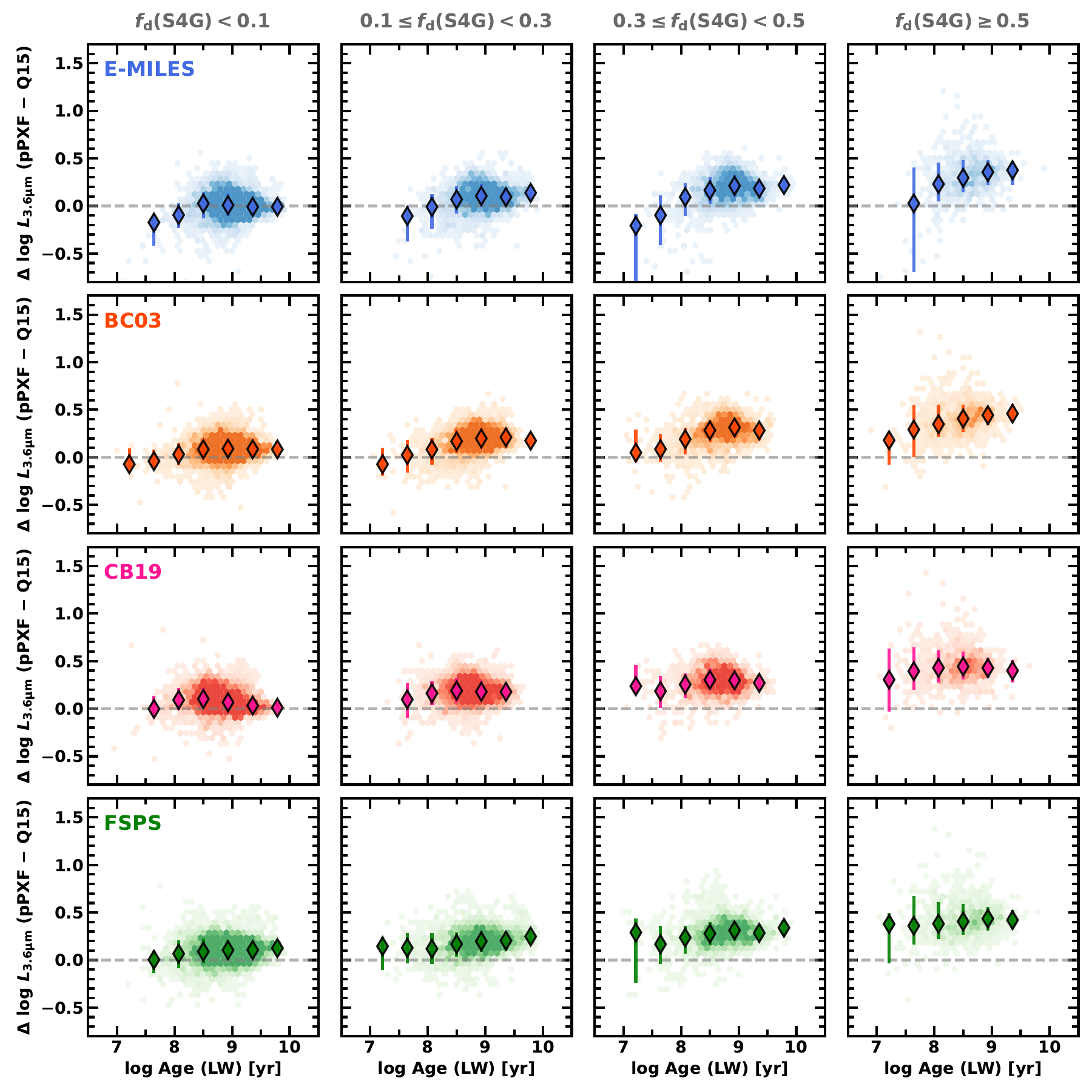}
\caption{
Same as {\color {blue} \textbf{Figure \ref{fig:dMass_vs_Age}}}, albeit for the difference in luminosity at IRAC $3.6~{\rm \mu m}$.
\label{fig:dFlux_vs_Age}}
\end{figure}

As shown in {\color{blue} \textbf{Figure \ref{fig:ML_vs_Age}}}, the mass-to-light ratios at $3.6~{\rm \mu m}$ derived from pPXF align well with the ${\rm S^{4}G}$ value for older stellar population (${\rm log~Age~(yr)>9.5}$) but are significantly lower for younger stellar populations.
To investigate the stellar mass differences with the Q15 method, we examined the influence of stellar ages, the non-stellar light fraction of the $3.6~{\rm \mu m}$ flux ($f_{\rm d}{\rm (S^{4}G)}$), and the choice of SPS models.
{\color{blue} \textbf{Figure \ref{fig:dMass_vs_Age}}} shows the distribution of mass differences as a function of stellar ages across different SPS models, with panels divided according to $f_{\rm d}({\rm S^{4}G})$ values.
As expected, trends with stellar age and SPS model are consistently observed.
The pPXF-derived stellar masses converge with the ${\rm S^{4}G}$ results for stellar populations older than ${\rm log~Age~(yr)}>9.5$, which is consistent with the underlying assumption of stellar ages ($2-12~{\rm Gyr}$) in Q15.
However, in regions with younger populations (${\rm log~Age~(yr)}<9$), our fiducial stellar masses are $\sim-0.5~{\rm dex}$ lower than ${\rm S^{4}G}$ masses, particularly for $f_{\rm d}{\rm (S^{4}G)}<0.1$ when using BC03 or CB19.
This reflects the age--$M_{\ast}/L_{\rm 3.6~\mu m}$ relation discussed in {\color{blue} \textbf{Section \ref{sec:age}}}, demonstrating that younger stellar populations are a key factor driving the differences between pPXF-derived masses and those from the Q15 method.

Across the horizontal bins of $f_{\rm d}{\rm (S^{4}G)}$, the relationships between age and mass difference exhibit consistent slopes for a single SPS model, but the level of mass difference shows a parallel increase as $f_{\rm d}{\rm (S^{4}G)}$ increases.
This effect of $f_{\rm d}{\rm (S^{4}G)}$ can be more clearly observed in {\color{blue} \textbf{Figure \ref{fig:dFlux_vs_Age}}}, which shows the difference in the stellar luminosity at $3.6~{\rm \mu m}$. 
In low-dust regions with $f_{\rm d}{\rm (S^{4}G)}<0.1$, there is no significant systematic bias in the stellar luminosity at $3.6~{\rm \mu m}$ for most SPHEREx-binned pixels.
This indicates that the predicted $3.6~{\rm \mu m}$ luminosity is almost unaffected by stellar ages and SPS models, unlike stellar masses.
However, in high-dust regions with $f_{\rm d}{\rm (S^{4}G)}\geq0.5$, the pPXF results show $0.4-0.5~{\rm dex}$ higher $3.6~{\rm \mu m}$ luminosity than those from ${\rm S^{4}G}$.
This effect contributes to the mass overestimation from pPXF in high-dust and old stellar regions across all SPS models. 
The possible origin of this discrepancy is that the dust component in these high-$f_{\rm d}{\rm (S^{4}G)}$ regions might be over-subtracted by the ICA method used in ${\rm S^{4}G}$.

\subsection{The WISE Color Scaling Relations}
\label{app:WISE}

\subsubsection{WISE Images}
\label{app:data_WISE}

In this appendix, the WISE data were used to compare the fiducial stellar masses obtained from full-spectrum fitting with those derived from the WISE scaling relations presented by J23.
The WISE images of 19 PHANGS-MUSE galaxies were obtained from the IRSA archive.
These auxiliary images were background-subtracted and reprojected following the procedures described in {\color{blue} \textbf{Appendix \ref{app:data_S4G}}}.
For the WISE bands, we applied the flux zeropoints for the Vega magnitude system, as provided in \citet{wri10}.

\subsubsection{Comparison with J23 Method}
\label{sec:comp_WISE}

The WISE color-based scaling relations proposed by J23 use the W1$-$W2 ([3.4]$-$[4.6]) and W1$-$W3 ([3.4]-[8.0]) colors, which are sensitive to stellar populations and star formation activity in galaxies.
Thus, the J23 scaling relations are designed to account for both stellar populations and dust contributions for estimating stellar masses of galaxies.
J23 presented two empirical relations for mass-to-light ratios as follows:
\begin{equation}
    {\rm log~}(M_{\ast}/L_{\rm W1})=A_{0}+A_{1}\times({\rm W1-W2})
    \label{eqn:w12}
\end{equation}
\begin{equation}
    {\rm log~}(M_{\ast}/L_{\rm W1})={\rm log~}[\phi_{\ast}\gamma^{(1+\alpha)}\exp{(-\gamma)}]
    \label{eqn:w13}
\end{equation}
where the coefficients in {\color{blue} \textbf{Equation \ref{eqn:w12}}} are $A_{0}=-0.376$ and $A_{1}=-1.053$. 
For {\color{blue} \textbf{Equation \ref{eqn:w13}}}, the parameters are $\phi_{\ast} = 0.454$ and $\alpha = -1.00$, and $\gamma$ is defined as $10^{[0.4\times(({\rm W1-W3})-({\rm W1-W3})^{\ast})]}$ with $({\rm W1-W3})^{\ast} = 4.690$.
Both the W1$-$W2 and W1$-$W3 colors are measured in the Vega magnitude system.
Applying the BC03 model in J23, these relations offer a simpler approach compared to full-spectrum fitting (this study) and Q15 method.
We applied these relations by measuring WISE fluxes and colors for all SPHEREx-binned pixels from the reprojected WISE images, as described in {\color{blue} \textbf{Appendix \ref{app:data_WISE}}}.
From these results, we evaluated the reliability of the WISE scaling relations by comparing them with our spectrum-derived stellar masses.

\begin{figure}
\centering
\includegraphics[width=1.0\textwidth]{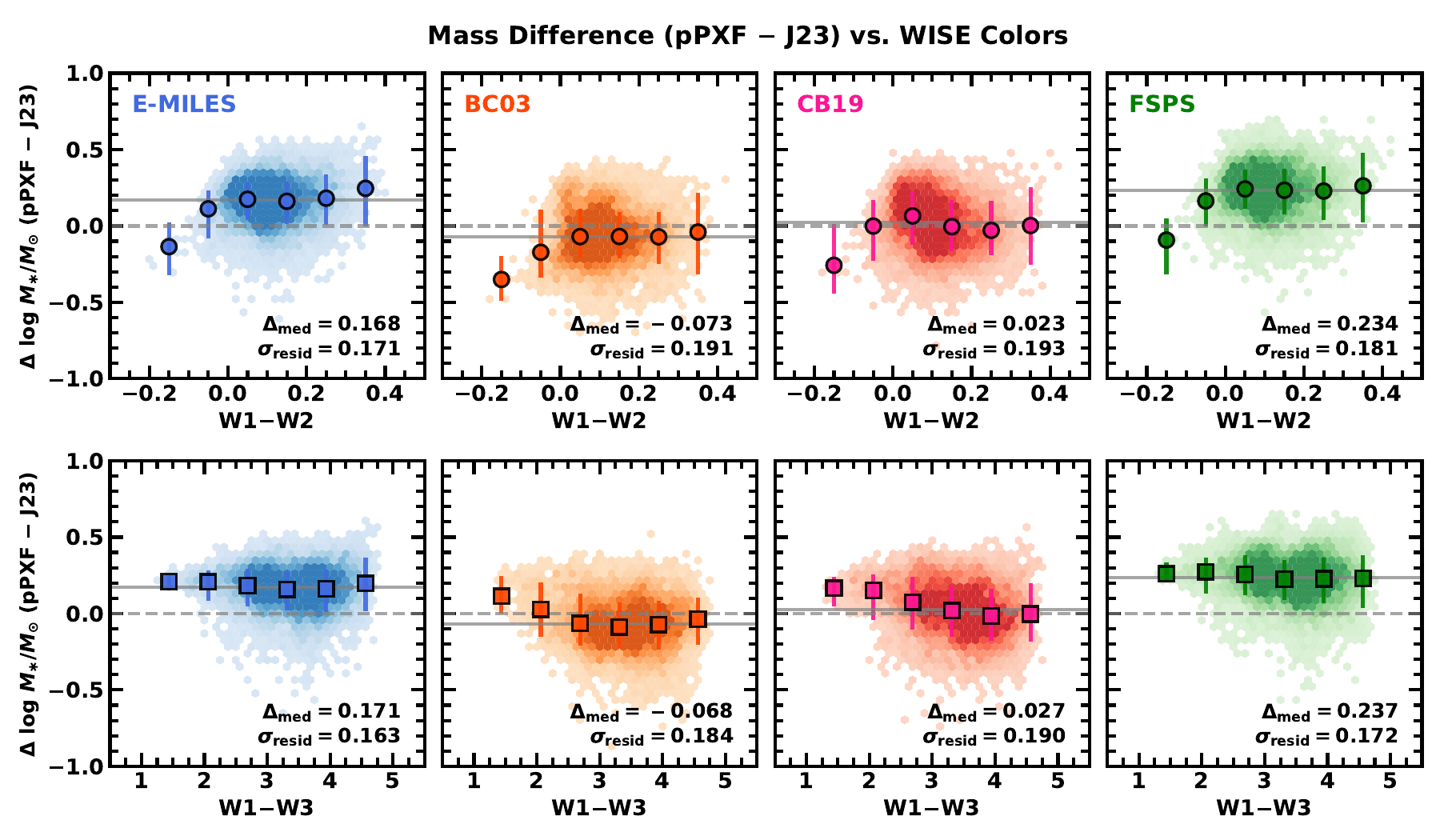}
\caption{
Comparison of stellar masses from this study (pPXF) with those derived from the WISE scaling relations (J23).
Each vertical column corresponds to a different SPS model: E-MILES (first column), BC03 (second column), CB19 (third column), and FSPS (fourth column).
The gray dashed line marks where the stellar masses from pPXF and J23 coincide.
Median mass differences, represented by gray solid lines, are shown across all color ranges, with circle and square symbols for each WISE color bin.
The median offsets and scatter (standard deviation of the residuals) are indicated in the bottom right corner of each panel.
\label{fig:wise2}}
\end{figure}

We compared our fiducial stellar masses obtained from full-spectrum fitting and those derived from the J23 relations, as shown in {\color{blue} \textbf{Figure \ref{fig:wise2}}}, which is divided horizontally by SPS models.
For the W1$-$W2 scaling relation, the overall trend of mass differences shows only weak dependencies on color, except in the blueward W1$-$W2 color range, where there is a clear underestimation of pPXF-derived masses with respect to J23 masses.
This may be attributed to the strong PAH $3.3~{\rm \mu m}$ emission in the W1 band and the absence of PAH contributions in the W2 band, which can boost the W1 band flux, resulting in bluer W1$-$W2 colors \citep{lee12}.
This effect leads to a systematic overestimation of J23 masses.
In contrast, the mass differences from the W1$-$W3 scaling relation show nearly no color dependencies, suggesting that W1$-$W3 colors effectively mitigate the effect of the PAH $3.3~{\rm \mu m}$ on stellar mass estimation.
This may be because both the W1 and W3 bands are influenced by PAH features at $3.3~{\rm \mu m}$ and $11.3~{\rm \mu m}$, respectively, which arise from C–-H bonds \citep{dul81, all85}, thereby alleviating contamination from PAH emissions in the W1$-$W3 color.
These comparisons reveal that using color information as demonstrated by J23 can effectively approximate the mass-to-light ratio in the NIR band, providing more robust stellar mass estimates than those derived from a constant mass-to-light ratio.

In both J23 relations, discrepancies among the SPS models persist, reflecting the relative offsets observed in {\color{blue} \textbf{Figure \ref{fig:violin}}}.
BC03 and CB19 provide the smallest median offsets for the W1$-$W2 and W1$-$W3 relations, while E-MILES and FSPS overestimate masses by $\sim0.2-0.3~{\rm dex}$ for both relations.
From these comparisons, we find that J23 stellar masses have systematic offsets ranging from $-0.07$ to $0.24~{\rm dex}$ and scatters of $\sim0.2~{\rm dex}$, relative to our spectrum-derived stellar masses.

\clearpage

\end{document}